\documentclass[12pt,preprint]{aastex}
\usepackage{graphicx}
\usepackage{color}
\usepackage{url}
\usepackage{CJKutf8}

\slugcomment{Accepted by \apj~on 2015 Sept 24}

\shorttitle{2015 D1}
\shortauthors{Hui et al.}

\RequirePackage{lineno}

\begin{document}

\title{Gone in a Blaze of Glory: the Demise of Comet C/2015 D1 (SOHO)}
\author{
\begin{CJK}{UTF8}{bsmi}
Man-To Hui (許文韜)$^{1}$, Quan-Zhi Ye (葉泉志)$^{2}$,
\end{CJK}
Matthew Knight$^{3}$, Karl Battams$^{4}$ 
and David Clark$^{2,5}$
}
\affil{$^1$Department of Earth, Planetary and Space Sciences,
UCLA, 
595 Charles Young Drive East, 
Los Angeles, CA 90095-1567\\
}
\affil{$^2$Department of Physics and Astronomy, 
The University of Western Ontario,
London, Ontario, N6A 3K7, Canada\\
}
\affil{$^3$
Lowell Observatory, 1400 W. Mars Hill Road, Flagstaff, AZ 86001\\
}
\affil{$^4$
Naval Research Laboratory - Code 7685, 4555 Overlook Avenue, 
SW, Washington, DC 20375\\
}
\affil{$^5$
Department of Earth Sciences, 
The University of Western Ontario,
London, Ontario, N6A 5B7, Canada\\
}
\email{pachacoti@ucla.edu}

\begin{abstract}

We present studies of C/2015 D1 (SOHO), the first sunskirting comet ever seen from ground stations over the past half century. The \textit{Solar and Heliospheric Observatory} (\textit{SOHO}) witnessed its peculiar light curve with a huge dip followed by a flareup around perihelion: the dip was likely caused by sublimation of olivines, directly evidenced by a coincident temporary disappearance of the tail. The flareup likely reflects a disintegration event, which we suggest was triggered by intense thermal stress established within the nucleus interior. Photometric data reveal an increasingly dusty coma, indicative of volatile depletion. A catastrophic mass loss rate of $\sim$10$^{5}$ kg s$^{-1}$ around perihelion was seen. Ground-based Xingming Observatory spotted the post-perihelion debris cloud. Our morphological simulations of post-perihelion images find newly released dust grains of size $a \gtrsim 10$ $\mu$m in radius, however, a temporal increase in $a_{\min}$ was also witnessed, possibly due to swift dispersions of smaller grains swept away by radiation forces without replenishment. Together with the fading profile of the light curve, a power law dust size distribution with index $\gamma = 3.2 \pm 0.1$ is derived. We detected no active remaining cometary nuclei over $\sim$0.1 km in radius in post-perihelion images acquired at Lowell Observatory. Applying radial non-gravitational parameter, $\mathcal{A}_{1} = \left(1.209 \pm 0.118 \right) \times 10^{-6}$ AU day$^{-2}$, from an isothermal water-ice sublimation model to the \textit{SOHO} astrometry significantly reduces residuals and sinusoidal trends in the orbit determination. The nucleus mass $\sim$10$^{8}$--10$^{9}$ kg, and the radius $\sim$50--150 m (bulk density $\rho_{\mathrm{d}} = 0.4$ g cm$^{-3}$ assumed) before the disintegration are deduced from the photometric data; consistent results were determined from the non-gravitational effects.

\end{abstract}


\section{INTRODUCTION}


Near-Sun comets offer valuable chances for studying the physical properties of cometary nuclei because they are prone to disintegration due to the proximity to the Sun around perihelion (e.g., Sekanina \& Chodas 2005). Nearly 3,000 of near-Sun comets have been discovered since the operation of the joint ESA/NASA \textit{Solar and Heliospheric Observatory} (\textit{SOHO}) in 1996, but the majority appeared dim and failed to achieve a high signal-to-noise ratio in the data. In many cases observations from \textit{SOHO} alone contain large astrometric and photometric uncertainties. Many questions yet remain unsettled. For example, what are the differences between near-Sun comets and other comets in terms of material compositions? Where did they originate and how did they evolve into near-Sun orbits? Thus, ground-based observations of near-Sun comets are highly encouraged, yet frequently hampered by the extremely harsh observational geometry. Consequently it is very rare that a near-Sun comet is observed from ground observatories, evidenced by merely five\footnote{We dismiss the case of C/2008 O1 (\textit{SOHO}), which was serendipitously detected in images of a total solar eclipse, after a search based upon \textit{SOHO} data (Pasachoff et al. 2009).} successful examples since the start of 21$^{\mathrm{st}}$ century, viz. C/2011 W3 (Lovejoy) (e.g., Lovejoy \& Williams 2011, Sekanina \& Chodas 2012, etc.), C/2012 E2 (SWAN)\footnote{Through private communications, we see that T. Lovejoy managed to obtain 3 astrometric positions from his images taken in strong dusk twilight on UT 2012 March 10.38, from Australia. However, his report remains largely unnoticed.}, C/2012 S1 (ISON) (e.g., Novski et al. 2012), 322P/1999 R1 (\textit{SOHO})\footnote{Published in Minor Planet Electronic Circular (MPEC) 2015-K84, \url{http://www.minorplanetcenter.net/mpec/K15/K15K84.html}.}, and the newly discovered comet C/2015 D1 (\textit{SOHO}) (hereafter 2015 D1).

2015 D1 was discovered by W. Boonplod in \textit{SOHO}'s Large Angle Spectrometric Coronagraph (LASCO) images from UT 2015 February 18 (Battams \& Knight 2015). Our orbital solution (discussed in Section \ref{orb_det}) to the LASCO astrometry confirms that 2015 D1 does not belong to the Kreutz sungrazing family (Kreutz 1888), or any of the known sunskirting families including the Meyer, Marsden and Kracht groups (c.f. Sekanina \& Chodas 2005 and citations therein). We searched for small bodies with similar orbits via the JPL Small-Body Database Search Engine, yet found nothing. The small perihelion distance, $q = 6.06~R_{\odot}$ ($1~R_{\odot} = 0.00465$ AU), is substantially greater than perihelia of the Kreutz sungrazing family ($q \lesssim 2~R_{\odot}$), and close to, but somewhat smaller than, perihelia of the Meyer, Marsden and Kracht sunskirting families (mean perihelia all have $q \gtrsim 7.7~R_{\odot}$). Knight \& Walsh (2013) discriminates near-Sun comets subjected to tidal fragmentation events as sungrazing comets. Although 2015 D1 apparently disrupted, the disruption was unlikely tidally driven due to the relatively large heliocentric distance. In this manner, we address 2015 D1 as a sunskirting comet, rather than a sungrazing one.

Given the classification, we realize that 2015 D1 is a unique sunskirting comet in that it is the brightest and the first sunskirting comet which was observed from the ground over the past half century\footnote{Successful ground observations of the other sunskirting comet 322P/\textit{SOHO} were made on UT 2015 May 22, later than observations of 2015 D1.}. This paper presents our photometric, morphological and orbital analysis of 2015 D1.

\section{Observations}

\subsection{\textit{SOHO}}
The \textit{SOHO} spacecraft is located around the L1 point of the Sun-Earth system. The LASCO instrument onboard consists of three coronagraphs, C1, C2 and C3. Only the C2 and C3 cameras observed 2015 D1. The C2 and C3 coronagraphs, externally occulted, have annular fields of view (FOV) of 1.5--6.0~$R_{\odot}$ and 3.7--30~$R_{\odot}$, respectively (Brueckner et al. 1995). Each instrument is equipped with a filter wheel, a polarizer wheel, a shutter, and a $1024 \times 1024$ pixel CCD with a pixel scale of 11\arcsec.9 pixel$^{-1}$ for C2, and 56\arcsec.1 pixel$^{-1}$ for C3. The synoptic C2 data are taken through an orange filter with bandpass $\sim$5400--6400 \AA, whereas the C3 observations are mainly made with a clear filter with bandpass $\sim$4000--8500 \AA. Other filters are used much less frequently, generally once per day, and these images have half resolution ($512 \times 512$). Each camera carries a polarizer wheel having polarizer positions of $-60\degr$, $0\degr$, and $+60\degr$, and takes polarization sequences 1--2 times per day.

2015 D1 was observed by C3 from UT 2015 February 18.0--21.8, mostly through the clear filter. The C2 camera also continuously monitored it around perihelion from UT 2015 February 19.6--19.9, all through the orange filter. Other available data include a few of C3 blue and orange filter images, and four triplets of polarized orange filter images. All the LASCO images were processed in a similar way as described in Knight et al. (2010) by use of SolarSoftWare (SSW) and SolarSoftWare DataBase (SSWDB)\footnote{SSW and SSWDB are both parts of the SolarSoft system, \url{http://www.lmsal.com/solarsoft/}.} in IDL.

\begin{figure}
\begin{center}
\includegraphics[scale=0.4]{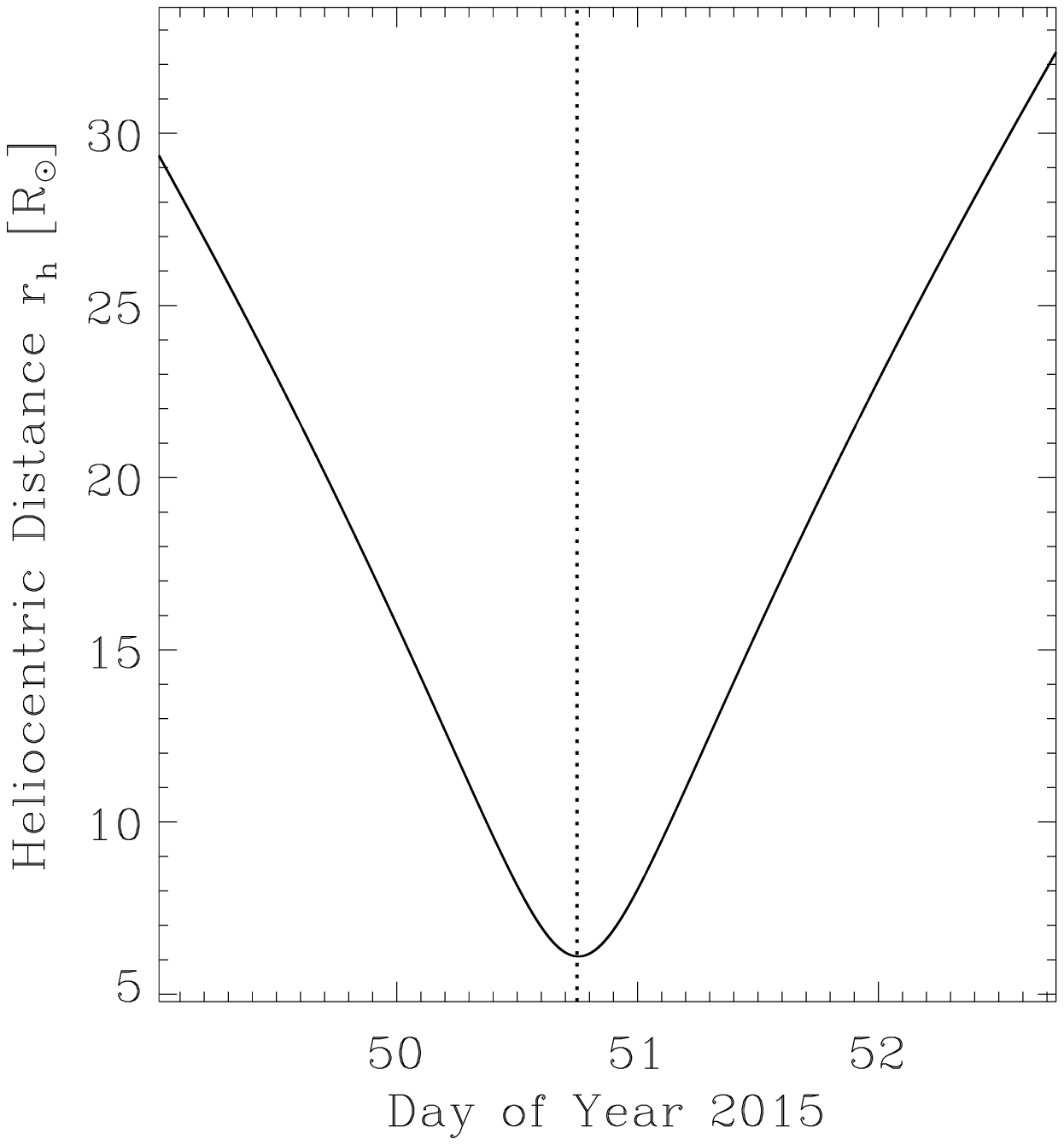}
\includegraphics[scale=0.4]{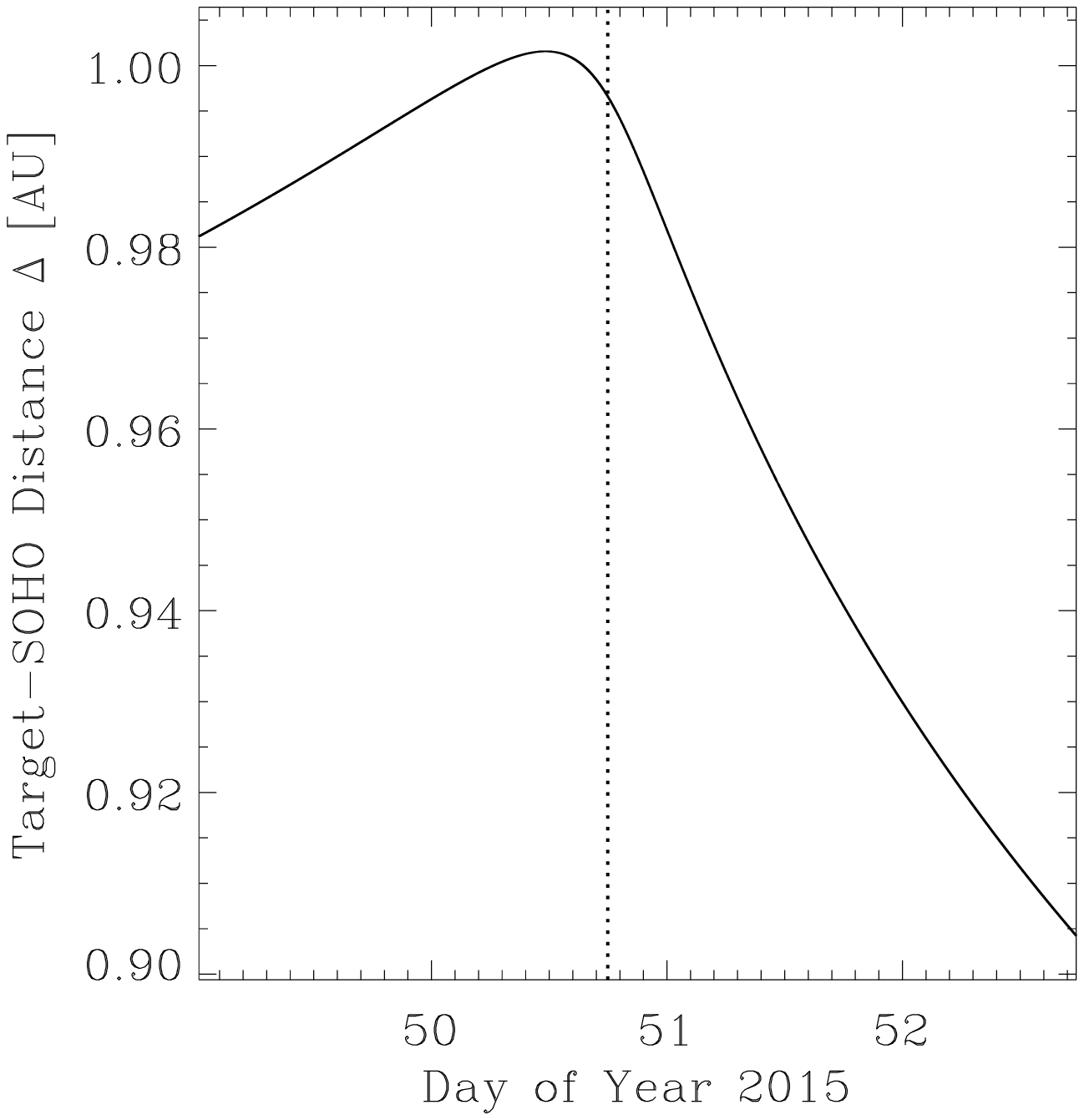}
\includegraphics[scale=0.4]{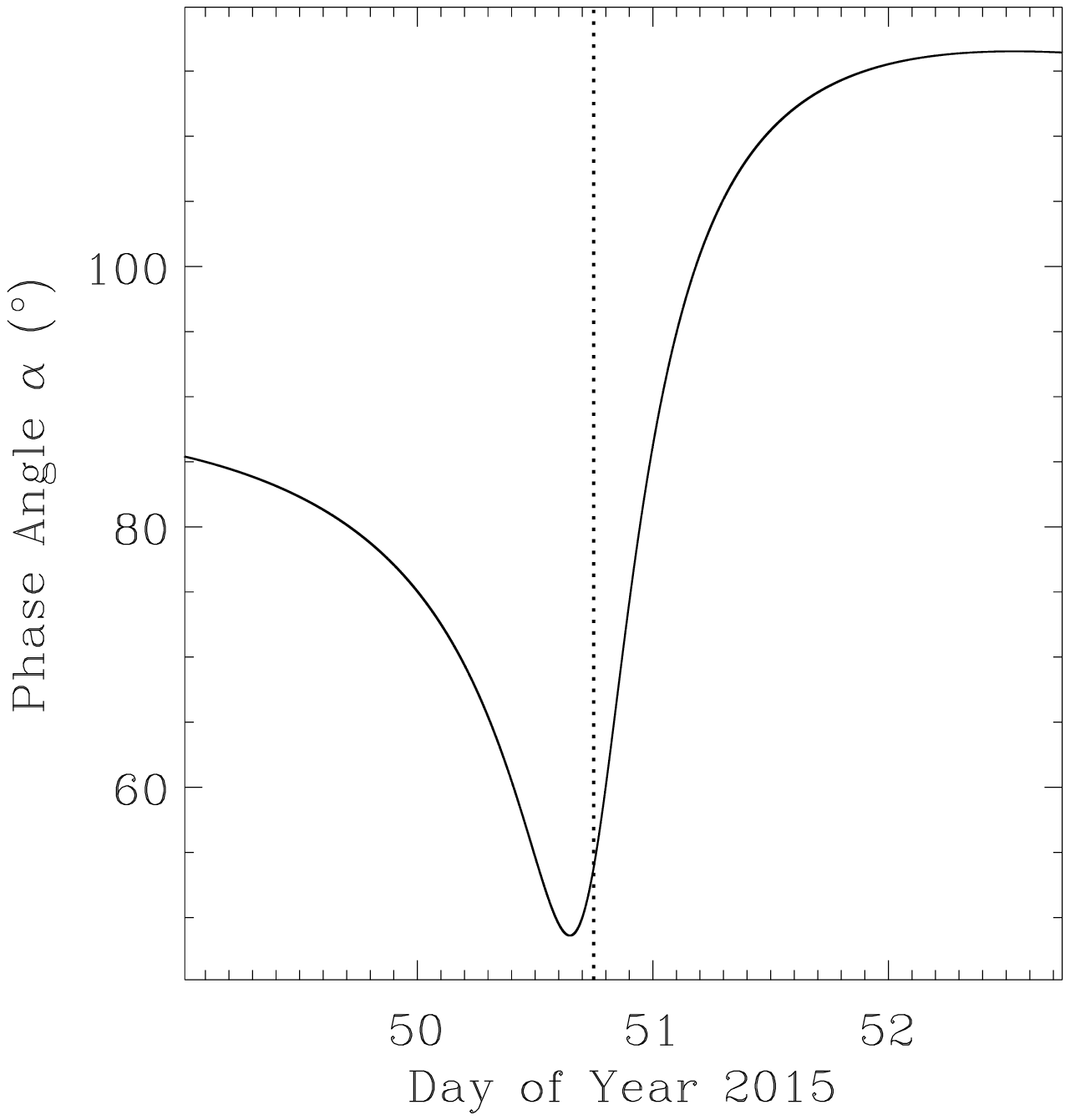}
\caption{
Observational geometry of 2015 D1 from \textit{SOHO}'s perspective during its transit in \textit{SOHO}'s FOV. The vertical dotted line in each panel marks the perihelion time $t_{\mathrm{P}}$ (UT 2015 February 19.75).
\label{fig:geometry}
}
\end{center}
\end{figure}

The observational geometry of 2015 D1 from \textit{SOHO} is illustrated by Figure \ref{fig:geometry}.

\subsection{Ground-Based Observations}

\subsubsection{Xingming Observatory}

We conducted post-perihelion observations of 2015 D1 on UT 2015 March 4, 8, 9 and 15 via the 10.6-cm f/5.0 refractor attached with an Apogee U16M $4096\times4096$ CCD through a photometric standard V-band filter as part of the Comet Search Program (CSP) of Xingming Observatory. The images have a square FOV $4\degr.0 \times 4\degr.0$, and a pixel scale of 3\arcsec.53 pixel$^{-1}$. Exposures of data taken from the first two nights were 60 s and 120 s in duration, whereas data from the last two nights had exposures of 120 s only. The image quality varied from night to night, generally $\sim$9\arcsec--10\arcsec~FWHM (Full Width Half Maximum). All the observation sessions were started from dusk, because of the small solar elongation. Images were first fully calibrated by subtracting bias and dark current, then were divided by flat-field frames, and finally were normalized by exposure times. The image sequence from each night was registered on field stars, and then was shifted following the motion of 2015 D1. Normalization of the sky background was then performed. Finally, the images were median co-added into a single frame. We are able to detect an enormous cigar-shaped nebulosity with its west tip within $\sim$5\arcmin~of the predicted positions either given by JPL HORIZONS or our orbit solutions regardless of including non-gravitational parameters. It had a dimension of $\sim1\degr \times 0.2$\degr, directed approximately east to west. The cloud appeared the most obvious on March 4, even discernible in individual frames, and the dimmest yet still sufficient for visual detection in the final stacked image from March 15.

\subsubsection{Lowell Observatory}

We attempted to recover 2015 D1 using Lowell Observatory’s 4.3-m Discovery Channel Telescope (DCT) on UT 2015 March 5. We used the Large Monolithic Imager (LMI), which has a FOV 12\arcmin.3 on a side and a 6.1K $\times$ 6.1K e2v CCD. Images were binned on chip $2\times2$, resulting in a pixel scale of 0\arcsec.24 pixel$^{-1}$. We obtained three 30 s images using the broadband Cousins R filter. Images were trailed at the comet’s rate of motion, and the pointing was determined by Lowell’s ephemeris calculator from the orbital solution published by the Minor Planet Center (MPEC 2015-D73). LMI was not scheduled to be used on this night, so these were the only three images obtained. We removed the bias and applied a flat field correction using images from 2015 February 25, which was the closest night to our observations on which science data were obtained. Observing conditions were poor because this was the first night following a series of winter storms, so atmospheric seeing was significantly worse than normal. Due to the necessity of acquiring images as early as possible following twilight, the default focus values were used, so the instrumental point spread function was likely suboptimal. 

On the same night we also imaged the comet’s field with Lowell Observatory’s 31-in (0.8-m) telescope. The 31-in has a 2K $\times$ 2K e2v CCD42-40 chip with a FOV 15\arcmin.7 on a side and a pixel scale of 0\arcsec.46. We obtained ten 30 s images with the Cousins R filter trailed at the comet’s rate. The bias was removed and the images were flat-fielded in the standard manner. 

Despite that 2015 D1 had a large 3$\sigma$ position uncertainty of $\sim$4\arcmin~during the observations, both of the FOVs are large enough to encompass the region. We visually searched both sets of images using several methods but did not find any evidence of the comet. We could detect field stars in DCT images to SDSS r magnitude of $\sim$20.0 (Ahn et al. 2012), and likely could have detected the comet to magnitude $\sim$21 despite the poor seeing since it would have been stationary while the stars were visibly trailed ($\sim$3\arcsec.5 or $\sim$15 pixels). We could detect field stars to an SDSS r magnitude of $\sim$19.0 in the 31-in images. This is likely the limiting magnitude for any comet non-detection with the 31-in since the stars did not appear significantly trailed due to the considerably worse seeing than on DCT.

Observation condition details from Xingming and Lowell Observatories are summarized in Table \ref{geometry}.

\section{Results}

\subsection{Photometry} 
\label{phot_ana}
Although the ground-based observations provided much better resolution than did the LASCO C2/C3 cameras, it is impossible to perform photometric measurements with them because of the extreme fuzziness of the debris cloud of 2015 D1 as well as the tremendous area it occupied. We only conducted aperture photometry of the comet in C2/C3 images by using packages in the IDL Astronomy User's Library (Landsman 1993). 

Apparent magnitudes were converted from the measured fluxes with zero-points of the LASCO images, which were calculated based upon trespassing field stars in LASCO data. Due to degradation effects of the LASCO detectors, the zero-points have changed slightly year by year. Only degradation influences upon C2 orange and C3 clear data have been examined exhaustively, as data for these filters is the most abundant. We cannot find out any detailed information about the changes in the zero-points of other filters. We adopted the temporal zero-point computed by Gard\`es et al. (2013) to calculate the C2 orange magnitude, and the zero-point by Lamy et al. (2013) to calculate the C3 clear magnitude. For other filters we used values given by Llebaria et al. (2006), Knight (2008) and citations therein, and further included an uncertainty of $\pm$0.05 mag in error estimates due to the unavailable temporal evolutions in the zero-points.

Because of the low spatial resolutions of the C2/C3 cameras, we used a fixed angular sized aperture, which allows direct comparison of our results to previous studies of near-Sun comets (e.g., Biesecker et al. 2002, Knight et al. 2010, Lamy et al. 2013). A circular aperture of radius 5 pixels (1\arcmin.0) was selected for full resolution $1024\times1024$ pixel C2 images and 3 pixels (2\arcmin.8) was selected for $1024\times1024$ pixel C3 images, to enclose the signal of 2015 D1, but at the same time to minimize contaminants from sky background as much as possible. Half resolution $512 \times 512$ pixel images had half sized apertures. The comet appeared overexposed in 14 C2 orange filtered images from UT 18:00--20:48, 2015 February 19, and hence we applied saturation corrections, developed by Knight et al. (2012). This likely still underestimates the total brightness slightly, but is much closer to the actual brightness. We estimate the uncertainties from the saturation correction at $<0.1$ mag and are systematic, e.g., nearby points have nearly identical saturation correction uncertainties. 

We converted apparent magnitudes $m_{V}$ into heliocentric magnitudes $H_{V}$, by normalizing the distance between \textit{SOHO} and 2015 D1 to $\Delta = 1$ AU, and correcting for the phase effect:
\begin{equation}
H_{V} \left(r_{\mathrm{h}} \right) = m_{V} \left(r_{\mathrm{h}}, \Delta, \alpha \right) - 5 \log \Delta + 2.5 \log \left[\phi \left(\alpha \right)\right] ,
\label{eq_mag}
\end{equation}
where $r_{\mathrm{h}}$ is the heliocentric distance, $\alpha$ is the phase angle, and the phase function $\phi \left(\alpha \right)$ is given by Marcus (2007) as
\begin{equation}
\phi\left(\alpha\right)=\frac{\delta_{90}}{1+\delta_{90}}\left[k\left(\frac{1+g_{\mathrm{f}}^{2}}{1+g_{\mathrm{f}}^{2}+2g_{\mathrm{f}}\cos\alpha}\right)^{3/2}+\left(1-k\right)\left(\frac{1+g_{\mathrm{b}}^{2}}{1+g_{\mathrm{b}}^{2}+2g_{\mathrm{b}}\cos\alpha}\right)^{3/2}+\frac{1}{\delta_{90}}\right] ,
\end{equation}
\noindent Here $\delta_{90}$ is the ratio of the dust-to-gas intensity observed at $\alpha=90$\degr, with $\delta_{90}=1$ for normal comets, $0 \le k \le 1$ is the partitioning coefficient between the forward- and back-scattering, and $g_{\mathrm{f}} > 0$ and $g_{\mathrm{b}} < 0$ are respectively the forward- and back-scattering asymmetry factors. Marcus (2007) suggested $k=0.95$, $g_{\mathrm{f}}=0.9$ and $g_{\mathrm{b}}=-0.6$ according to observations of six comets. The model has been applied widely in recent works regarding observations of near-Sun comets, e.g., Knight et al. (2010), Knight \& Battams (2014), etc. Although Li \& Jewitt (2015) found slightly different parameters for C/2010 X1 (Elenin), it has minimal effects with corrections always $<0.3$ mag to 2015 D1's data and does not meaningfully alter the light curve shape. Several of the comets examined by Marcus (2007) have perihelia considerably smaller than that of C/2010 X1 and thus the results are likely more comparable to 2015 D1. Therefore, we follow the suggested parameters by Marcus (2007). 

We assign $\delta_{90} = 1.0$ for C3 clear filter, $\delta_{90} = 0.39$ for C2/C3 orange filters, and $\delta_{90} = 10$ for C3 blue filter, from analysis of 2015 D1's color (see Section \ref{disc_color} for details). Since the comet did not experienced strong forward- or back-scattering effects ($\alpha \sim$50--115\degr), its phase function is relatively flat, so the exact choice of $\delta_{90}$ always has corrections $< 0.2$ mag. The general shape of 2015 D1's light curve would not have been affected by the phase function profoundly.

\begin{figure}
  \centering
  \begin{tabular}[b]{@{}p{1\textwidth}@{}}
    \centering\includegraphics[scale=0.6]{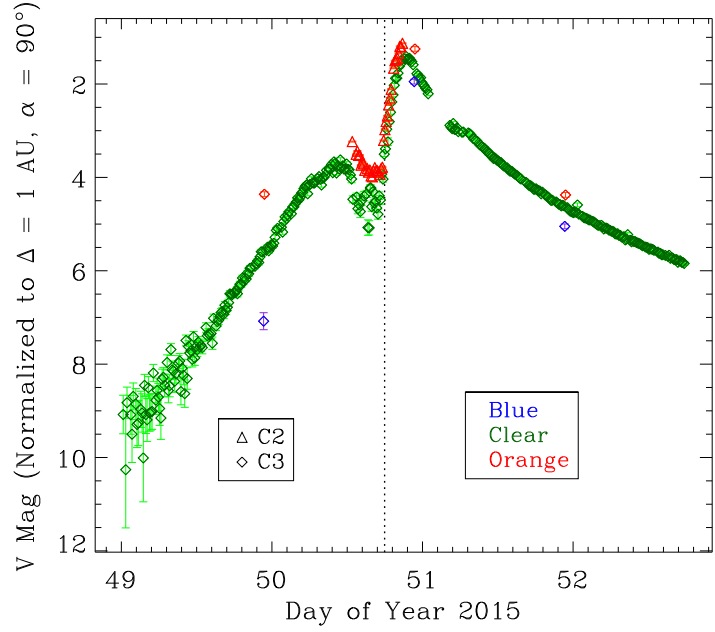} \\
    \centering\small (a)
  \end{tabular}%
  \quad
  \begin{tabular}[b]{@{}p{1\textwidth}@{}}
    \centering\includegraphics[scale=0.6]{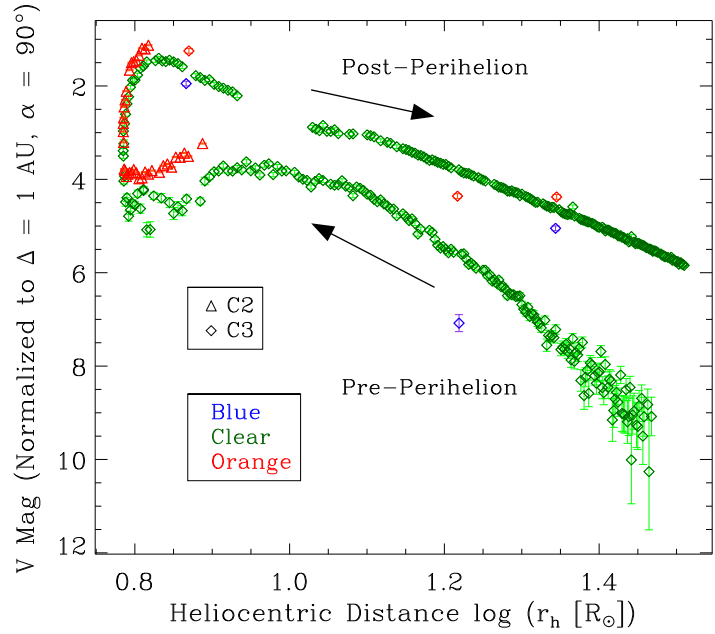} \\
    \centering\small (b)
  \end{tabular}
  \caption{
V band heliocentric magnitude of 2015 D1 observed by \textit{SOHO}/LASCO as (a) a function of time and (b) a function of heliocentric distance. Point symbols correspond to telescopes and points are color coded according to filters, as shown in the legend. The upper panel labels perihelion by a vertical dotted line. The two arrows in the lower panel sketch the direction of the comet's evolution.
\label{fig:lc_2015D1}
  }
\end{figure}

The resulting light curve of 2015 D1 is shown in Figure \ref{fig:lc_2015D1}a. The comet steadily brightened from the beginning at $H_{V} \sim 9$, until UT 2015 February 19.4 (denoted as $\Delta t = t - t_{\mathrm{P}} \sim -8$ hrs, where $t$ is observation epoch, and $t_{\mathrm{P}}$ is the perihelion time of the comet, UT 2015 February 19.75), when it apparently faded by $\sim$1 mag in $\sim$7 hrs, followed by a drastic surge in its brightness to $H_{V} \simeq 1.5$ through the clear filter, in $\sim$5 hrs. Post-perihelion witnessed a decline in its brightness. The comet was then obstructed by the pylon of the coronagraph for 3.6 hrs. Starting from $\Delta t \sim +0.6$ day (UT 2015 February 20.3 or DOY $\sim 51.3$) it dimmed smoothly on the way out of C3's FOV.

Figure \ref{fig:lc_2015D1}b shows $H_{V}$ as a function of $r_{\mathrm{h}}$. We can see that the post-perihelion brightness was consistently brighter than the pre-perihelion brightness at the same heliocentric distance, by $\gtrsim 1.5$ mag. The pre-perihelion brightening at $r_{\mathrm{h}} \gtrsim 13~R_{\odot}$, $\propto r_{\mathrm{h}}^{-5.5}$, was steeper than the post-perihelion fading at the same range, $\propto r_{\mathrm{h}}^{-2.8}$. A turnover point in the inbound leg at $r_{\mathrm{h}} \sim 13~R_{\odot}$ is noticed, where the brightening slowed down to $\propto r_{\mathrm{h}}^{-0.8}$. The second turnover point in the inbound leg occurred at $r_{\mathrm{h}} \sim 8~R_{\odot}$, after which the comet faded despite continuing to approach the Sun. Then the flareup took place around perihelion at $r_{\mathrm{h}} \sim 6~R_{\odot}$, and subsided at outbound $r_{\mathrm{h}} \simeq 6.7~R_{\odot}$. Similar light curves have been found amongst some of the Kreutz group comets (e.g., Knight et al. 2010). Starting from $r_{\mathrm{h}} \sim 13~R_{\odot}$ in the outbound leg, the comet faded steadily toward the end of the LASCO observation. The two respectively inbound and outbound turnover points at $r_{\mathrm{h}} \sim 13~R_{\odot}$ are very similar to those of Kreutz sungrazing comets, which are believed to be related to sublimation of olivines (e.g., Kimura et al. 2002).

As shown in Figure \ref{fig:color_2015D1}, the color of 2015 D1 was initially distinctly different from the color of the Sun, yet eventually evolved towards it, indicating that the coma became increasingly dusty. We think that this was due to depletion of sodium, which emitted strongly at the beginning of the LASCO observation, and faded out gradually. The comet had mean color indices Clear $-$ Orange $= +0.6$, and Clear $-$ Blue $= -0.7$ (see Section \ref{disc_color}). 

Four triplets of LASCO polarizer observations of 2015 D1 are available, however, they contain large uncertainties and the number of data points is too small, so the result will not be presented.

\subsection{Morphology}
\label{morph}

Using \textit{SOHO} and ground observations, we identified five stages in the evolution of 2015 D1:

\begin{enumerate}
 \item From discovery to $\Delta t \sim -8$ hr (UT 2015 February 18.0--19.4): the comet, initially almost stellar, was trailed by a developing faint tail as it brightened. It was similar to comparably bright Kreutz sungrazing comets at similar heliocentric distances (see Figure \ref{fig:img_soho}a\footnote{Note that the time of Figure \ref{fig:img_soho}a taken is not within this stage. However, this is the best image which shows the existence of the pre-perihelion tail, although it started to weaken.}).
 \item From $-8~\mathrm{hrs} \lesssim \Delta t \lesssim 0$ (UT 2015 February 19.4--19.7): the tail weakened and disappeared, whereby the comet became completely stellar (see Figure \ref{fig:img_soho}b).
 \item Within $0 \lesssim \Delta t \lesssim +1$ day (UT 2015 February 19.7--20.8): the comet developed a new tail, which was much more prominent than the pre-perihelion tail, at the same time that it brightened by about $\sim$3 mag (Figure \ref{fig:lc_2015D1}a). The optocentric region remained tight (see Figure \ref{fig:img_soho}c).
 \item From $\Delta t \sim +1$ day until the departure from \textit{SOHO}'s FOV (UT 2015 February 20.8--21.7): the comet maintained its tail, but the optocentric region appeared elongated (Figure \ref{fig:img_soho}d), reminiscent of some notable comet disintegration events such as C/1999 S4 (LINEAR) and C/2012 S1 (ISON) (e.g., Weaver et al. 2001, Knight \& Battams 2014).
 \item A week or more after the perihelion passage: multiple ground-based observers reported a nebulous cigar-shape object near the nominal position of 2015 D1 (e.g., Ma\v{s}ek et al. 2015). From these images, we identified no clear central condensation or a nucleus. The object dissipated rapidly as time went by. To our knowledge, no successful observation has been reported after mid-March.
\end{enumerate}

We interpret the physical evolution of 2015 D1 as follows. During its pre-perihelion phase, the comet behaved like a typical comet, with ongoing activity producing a dust tail (Figure \ref{fig:img_soho}a). But close to perihelion, dust began to sublimate faster than it was replenished and the tail disappeared (Figure \ref{fig:img_soho}b). The rapid brightening around perihelion and the subsequent development of a new tail seemed to indicate a sudden surge in activity of the comet. Considering the signs of nucleus disintegration depicted in subsequent images, it is apparent that such dramatic change of morphology reflects a catastrophic event experienced by the nucleus. The low spatial resolutions of \textit{SOHO} images hamper us from immediately looking into details of the disintegration, but it appears that the time from the flareup to the ultimate disruption of the nucleus took no more than 1 day (see Section \ref{sim_morph}). Generally speaking, the morphological evolution of 2015 D1 carries many similarities to that of Kreutz sungrazing comet C/2011 W3 (Lovejoy) (Sekanina \& Chodas 2012).

  \subsection{Orbital Determination and Non-Gravitational Effect}
  \label{orb_det}
 
 We only used the \textit{SOHO} astrometric data for orbit determination. 2015 D1 appeared too diffuse in ground-based observations, in spite of much better resolutions. Without a central condensation it is impossible to conduct astrometric measurements from these data.
  
 \textit{SOHO} astrometric measurements were recorded in custom software operated in IDL. The basic procedure was to manually select the optocenter of the comet and then allow the software to automatically calculate centroids on the 25 closest stars to the comet. This process occurred for every image in which the comet was visible. In the case of LASCO C2 there were not always 25 stars available, and thus as many as possible were recorded. LASCO C3 always has many more than 25 stars available. The limit of 25 stars has been selected as an optimum number based on computations of \textit{SOHO}-discovered comets in the early part of the \textit{SOHO} mission. All object locations were recorded at the sub-pixel level and passed to an implemented version of the Charon algorithm\footnote{\url{http://www.projectpluto.com/charon.htm}.}, which reduced the observations to a standard MPC format\footnote{\url{http://www.minorplanetcenter.net/iau/info/ObsFormat.html}}.
 
 We used EXORB\footnote{EXORB, a companion package of SOLEX, is an orbit determination program written by A. Vitagliano, available at \url{http://chemistry.unina.it/~alvitagl/solex/}.} to determine 2015 D1's orbit. Perturbations by all the eight planets, Pluto, and the three most massive asteroids, Ceres, Vesta and Pallas, are included in the computation using DE406 ephemerides, although they have basically no influence on solutions. Different weightings were assigned to the observations according to pixel scales. We filtered out 10 data points with residuals $\ge 50\arcsec$ as a cutoff. The remaining 412 observations all satisfy the residual threshold regardless of considering non-gravitational effects.
   
 The arc covered by the \textit{SOHO} observation was larger than any other \textit{SOHO}-discovered comets so we treated its eccentricity $e$ as one of the free parameters to be solved. We found that including the solving of non-gravitational parameters $\mathcal{A}_{j} \mbox{ ($j = 1, 2, 3$)}$, which are defined in Marsden et al. (1973)\footnote{Marsden et al. (1973) denoted the non-gravitational parameters by $A_{j} \mbox{ ($j = 1, 2, 3$)}$. We feel it necessary to change the symbol a little bit so as to avoid potential ambiguity with the used letter $A_{\mathrm{p}}$, for albedo.} from an isothermal water-ice sublimation model, significantly reduces the sinusoidal trends in astrometric residuals (the differences between the observed and calculated positions, a.k.a. O$-$C residuals, see Figure \ref{fig:orb_resid}). The trends are irrelevant to the selection of the astrometric data, in that filtering further more different sets of measurements to stricter residual thresholds or the otherwise, or removing data points apparently close to the edge of unblocked regions where diffraction by the occulter and the pylon of LASCO might take place do not alter the trend whatsoever. Other factors including infrequent resets of SOHO's onboard spacecraft clock and potential position errors of the spacecraft have been fully ruled out. We thus conclude that the residuals are authentic.
 
  We also found that solving the radial component $\mathcal{A}_{1}$ alone reduces the root-mean-square (RMS) of the best fit most noticeably, from $\pm$13\arcsec.37~to $\pm$10\arcsec.44. We obtained $\delta\mathcal{A}_{2}>1$ and $\delta\mathcal{A}_{3}=0.31$, the relative errors of $\mathcal{A}_{2}$ and $\mathcal{A}_{3}$ respectively, significantly larger than $\delta\mathcal{A}_{1}=0.09$. Taking into account the poor spatial resolutions of LASCO images, we solved  $\mathcal{A}_{1} = \left(+1.209 \pm 0.118 \right) \times 10^{-6}$ AU day$^{-2}$ only and simply assigned $\mathcal{A}_{j} = 0$ for $j \ne 1$. Similarly, JPL HORIZONS gives $\mathcal{A}_{1} = \left(+1.250 \pm 0.097 \right) \times 10^{-6}$ AU day$^{-2}$ with $\mathcal{A}_{j} = 0$ for $j \ne 1$ assumed\footnote{Retrieved on 2015 March 24.}. Different weightings and the number of observations filtered by JPL HORIZONS may account for the different values.
     
 Also tested was the forsterite sublimation model by Sekanina \& Kracht (2015). But we do not prefer that it was the mechanism responsible for the non-gravitational effect experienced by the comet, therefore, we did not apply it for the orbit determination (see Section \ref{disc_mloss} for details). 
 
 Our solutions to the orbital elements of 2015 D1 are listed in Table \ref{orbit}.

 \section{Discussion}
 
 \subsection{Search for Potential Pre-discovery Data}
 \label{pre_at}
 
 We investigated whether serendipitous imaging of 2015 D1 may have occurred. We applied EXORB to perform multiple iterations of Monte Carlo runs, based upon the random exclusion of a stochastically varying fraction (between 30 -- 70\%) of the \textit{SOHO} astrometric data, whereby 352, an arbitrary number, Monte Carlo clones of orbital elements of the comet were generated. Techniques documented in Clark (2010) were then applied to search for serendipitous pre-discovery imaging of the comet. Using the online Canadian Astronomy Data Centre Telescope Products (Gwyn et al. 2012) and the Minor Planet Center Sky Coverage Pointing Data dataset, over 600,000 archival images were considered from Canada-France-Hawaii Telescope (CFHT), Panoramic Survey Telescope and Rapid Response System (Pan-STARRS), Lincoln Near-Earth Asteroid Research (LINEAR), Spacewatch, Catalina, and Mount Lemmon, and approximately 50 smaller surveys, as well as visible and infrared images from spacecraft, including \textit{Wide-field Infrared Survey Explorer} (\textit{WISE}), and its asteroid-hunting portion, NEOWISE, \textit{Hubble Space Telescope}'s Wide Field Planetary Camera 2 (WFPC2) and Wide Field Camera 3 (WFC3) catalogues. 1,000 probability clones were generated and distributed over a spatial volume consistent with observational errors. Unfortunately no recent serendipitous image of any of these clones was identified. The MPC Sky Coverage Pointing Data dataset list three images from Pan-STARRS dated 2014 April 09 and 10 encompassing the position of 2015 D1, however the comet would have been too dim to be detected at these early dates, when $r_{\mathrm{h}} \gtrsim 5$ AU.

 \subsection{Pre-Perihelion Dip in Light Curve}
 \label{swings_eff}
 The \textit{SOHO} observations showed a dip in the light curve starting $\sim$8 hrs before perihelion (Figure \ref{fig:lc_2015D1}a). As sodium emission probably contributes significantly to the total brightness at small $r_{\mathrm{h}}$, one may question whether the dip was caused by the Swings effect (Swings 1941), i.e., temporal variation in intensity of cometary emission lines coincident with Fraunhofer lines due to Doppler shift. Assuming the entire gas emission was dominated by sodium, we use the following equation
 \begin{equation}
 H_{V, \mathrm{gas}} = H_{V} + 2.5 \log \left[ \left( 1 + \delta_{90} \right) \mathcal{G} \left( \dot{r}_{\mathrm{h}} \right) \right],
 \label{eq_heliomag_gas}
 \end{equation}
 \noindent where $\mathcal{G}$ is the normalized g-factor for sodium due to the Swings effect as a function of heliocentric radial speed $\dot{r}_{\mathrm{h}}$, to examine if the dip would be largely removed. We extracted a g-factor from Figure 2 in Watanabe et al. (2003) and normalized it to large $\dot{r}_{\mathrm{h}}$. The normalized g-factor has $\mathcal{G} = 1$ for $\dot{r}_{\mathrm{h}} \gtrsim 80$ km s$^{-1}$ and $\mathcal{G} = 0.05$ at perihelion.
 
 We find that not only does Equation (\ref{eq_heliomag_gas}) fail to remove the dip in the light curve, but also artificially creates a sharp brightening spike at perihelion. Despite the scatter in the data around the bottom of the dip, the maximum dimming $\sim$1 hr prior to the perihelion can still be recognized. It is extremely unlikely that perihelion is off by the $\sim$1 hr that would be necessary to reconcile it with the Swings effect since this would be more than two orders of magnitude larger than the 1$\sigma$ uncertainty in perihelion (see Table \ref{orbit}). Moreover, were it due to the Swings effect, the dip should occur much more abruptly such that a much sharper valley would be formed. More evidence which can help exclude the possibility of the Swings effect is that it would shrink the clear $-$ orange magnitude difference centering about the dip, which was not seen whatsoever. Therefore the Swings effect is unlikely to be relevant to the formation of the dip. 
 
 Likewise, we do not feel that instrumental vignetting can account for the observed dip. Admittedly, the minimum of the dip observed by LASCO C3 took place almost exactly when the vignetting is locally highest so any inappropriate vignetting correction may cause some effects. However, the C2 vignetting is small and relatively constant during the time of the cometary light curve dip. LASCO's vignetting functions are well-established as part of the instrument's calibrations, and accordingly we see no impact on our measurements or results that may arise from this correction. We therefore reject the possibility of the LASCO vignetting as the reason for the dip; intrinsic activity of 2015 D1 is more likely to be the cause.

We notice that the onset of the dip occurred at $r_{\mathrm{h}} \sim 8~R_{\odot}$, following a mild turnover at $r_{\mathrm{h}} \sim 13~R_{\odot}$, consistent with the light curves of Kreutz sungrazing comets. It is thus possibly analogous to the turnover in pre-perihelion brightness of the Kreutz sungrazing comets (e.g., Biesecker et al. 2002, Knight et al. 2010), which is believed to be correlated with the onset of sublimation of olivines (e.g., Kimura et al. 2002). The disappearance of the tail around perihelion also lends strong support to this idea.

 \subsection{Color}
 \label{disc_color}
\begin{figure}
  \centering
  \begin{tabular}[b]{@{}p{1\textwidth}@{}}
    \centering\includegraphics[scale=0.6]{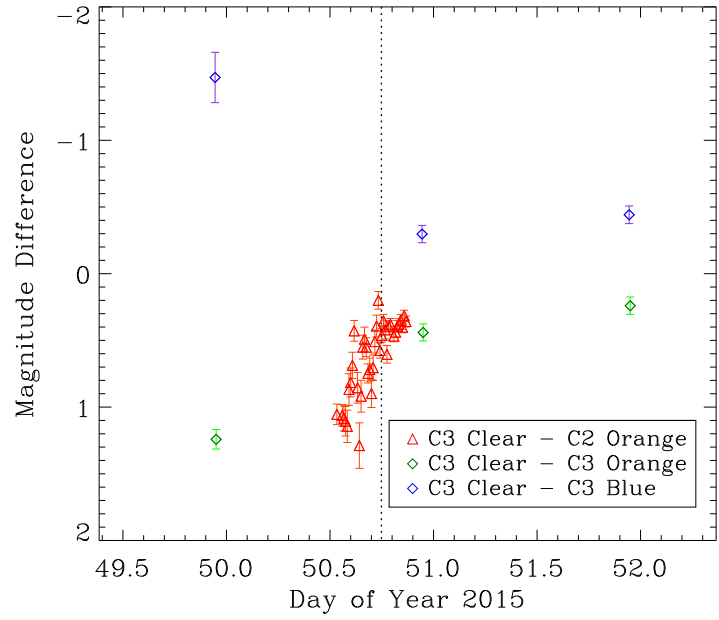} \\
    \centering\small(a)
  \end{tabular}%
  \quad
  \begin{tabular}[b]{@{}p{1\textwidth}@{}}
    \centering\includegraphics[scale=0.6]{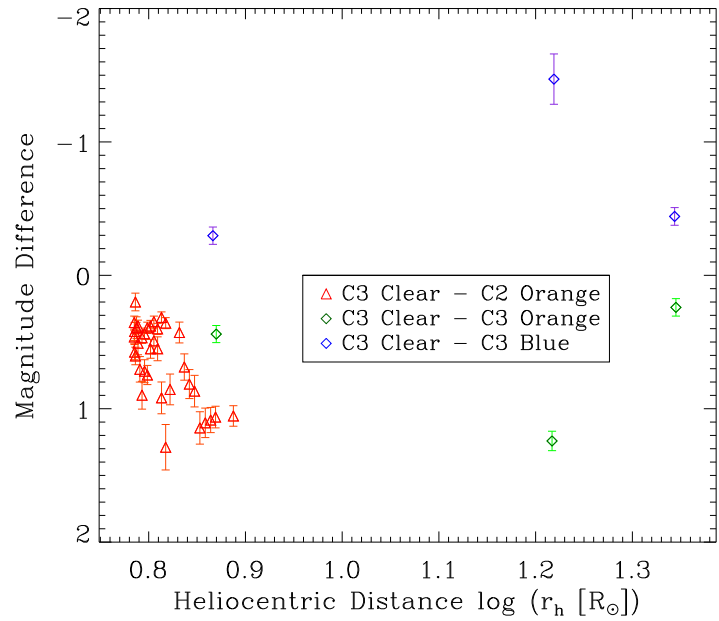} \\
    \centering\small (b)
  \end{tabular}
  \caption{
Color of 2015 D1 observed by \textit{SOHO}/LASCO as (a) a function of time and (b) a function of heliocentric distance. The upper panel shows the perihelion moment by a vertical dotted line.
  \label{fig:color_2015D1}
  }
\end{figure}

\begin{figure}
\epsscale{1.0}
\begin{center}
\plotone{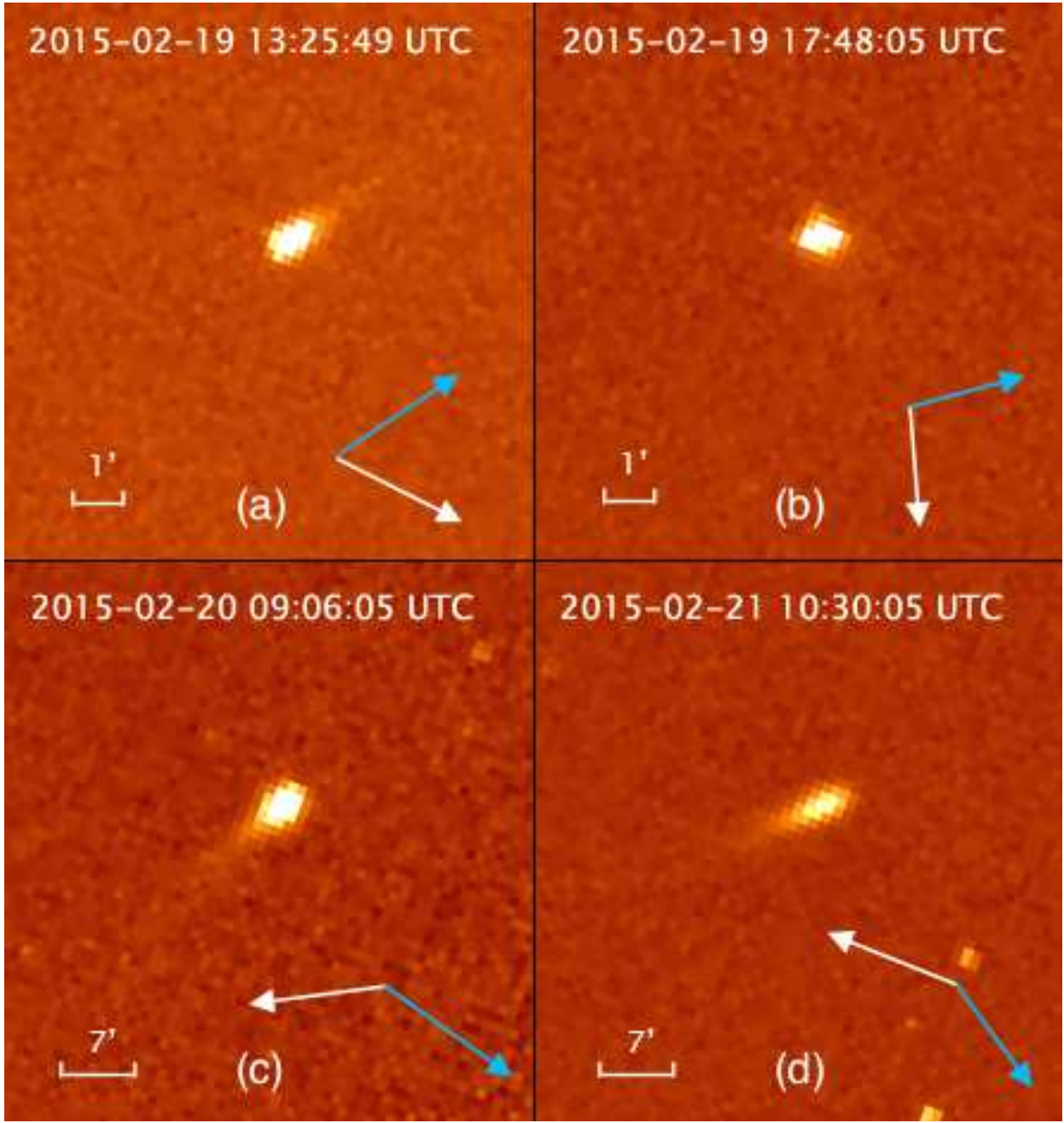}
\caption{Morphological evolution of 2015 D1 observed by \textit{SOHO}. The top two panels, (a) \& (b), are LASCO C2 images and the bottom two, (c) \& (d), are LASCO C3. In each panel, North is to the top and East to the left. The blue arrows point to the projected negative heliocentric velocity vector, and the white arrows point to the projected anti-solar direction. 
\label{fig:img_soho}
} 
\end{center} 
\end{figure}

\begin{figure}
  \centering
  \begin{tabular}[b]{@{}p{0.3\textwidth}@{}}
    \centering\includegraphics[width=0.32\textwidth]{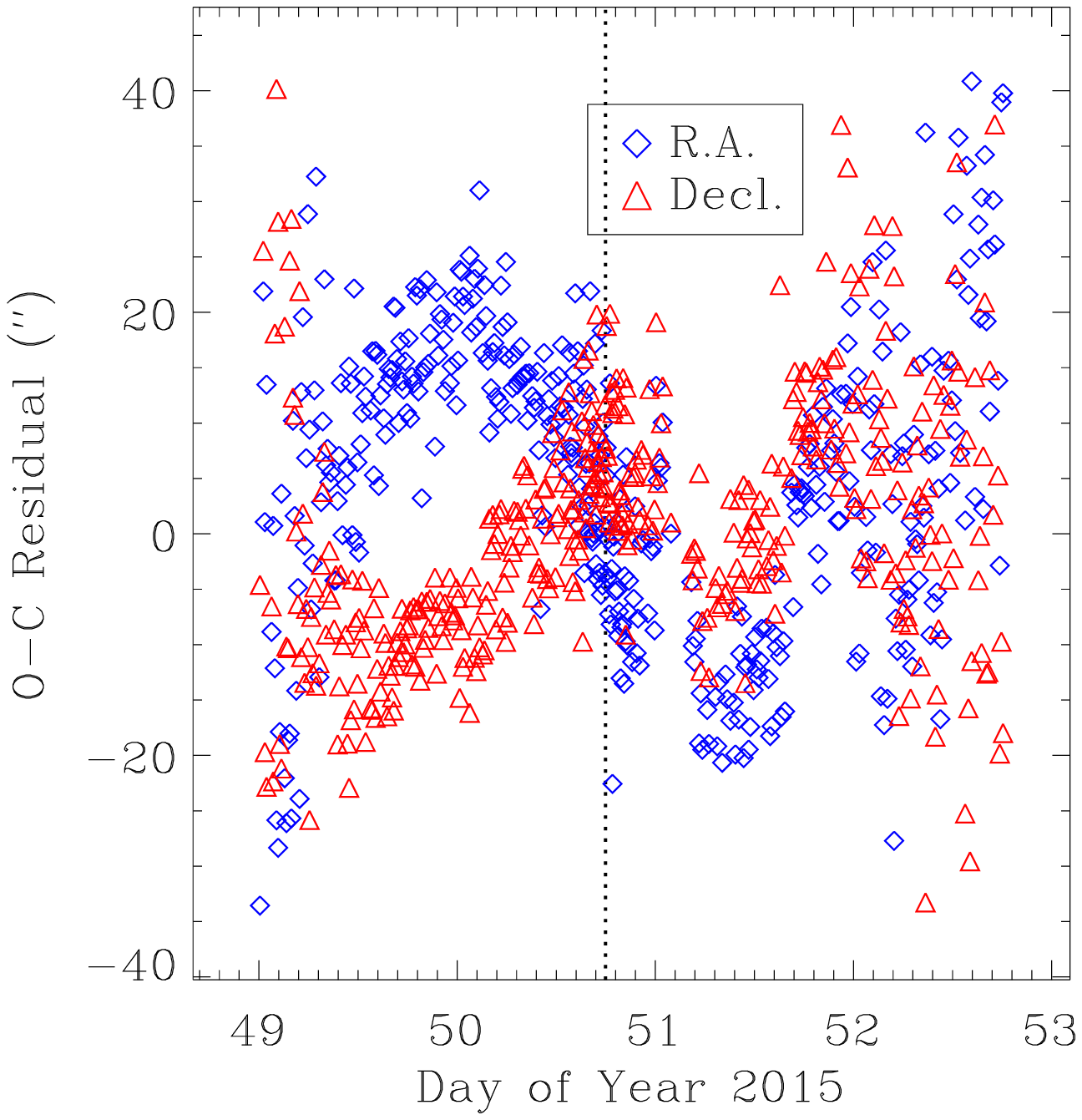} \\
    \centering\small ~~(a)
  \end{tabular}%
  \quad
  \begin{tabular}[b]{@{}p{0.3\textwidth}@{}}
    \centering\includegraphics[width=0.32\textwidth]{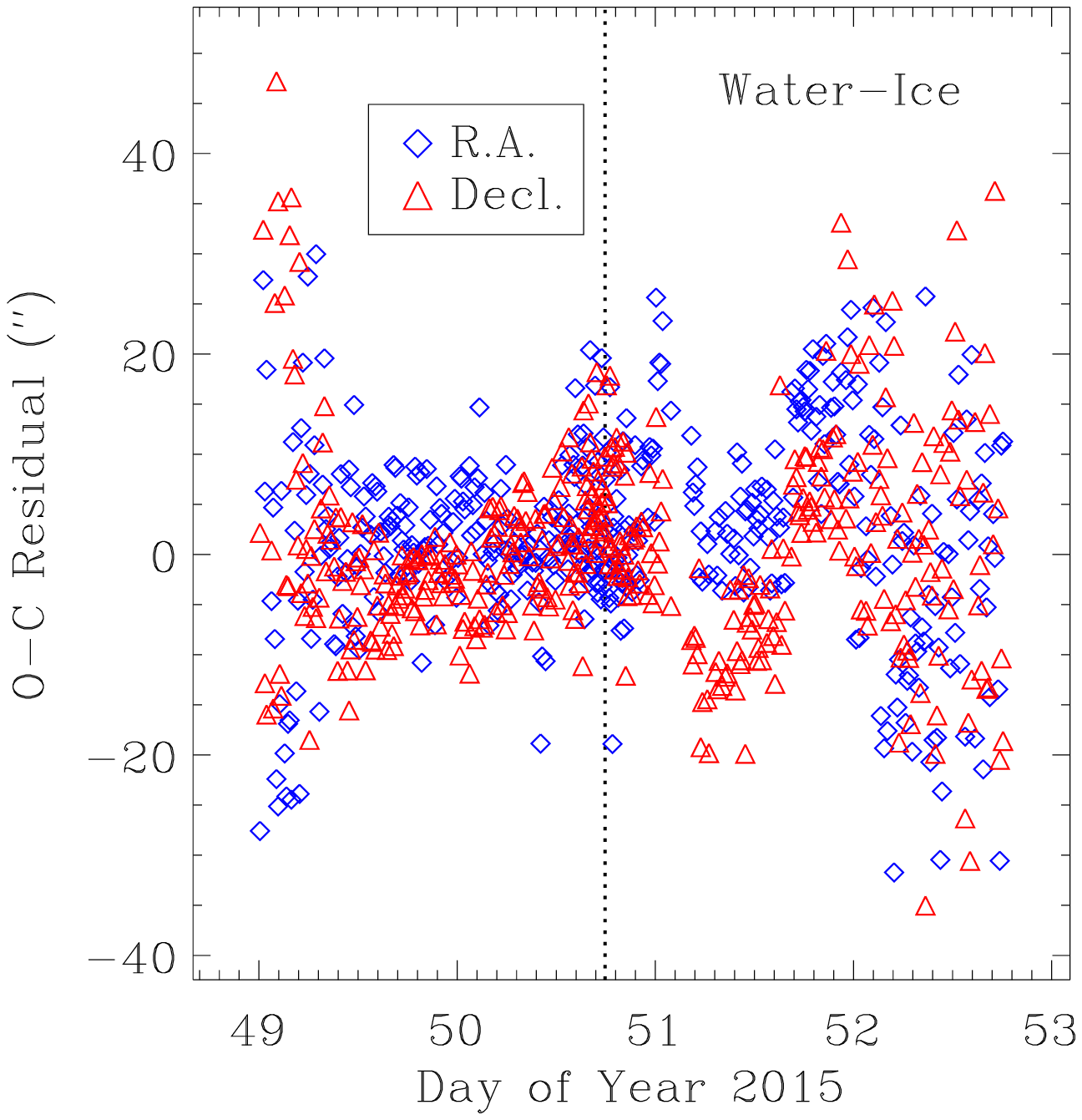} \\
    \centering\small ~~(b)
  \end{tabular}
  \quad
  \begin{tabular}[b]{@{}p{0.3\textwidth}@{}}
    \centering\includegraphics[width=0.32\textwidth]{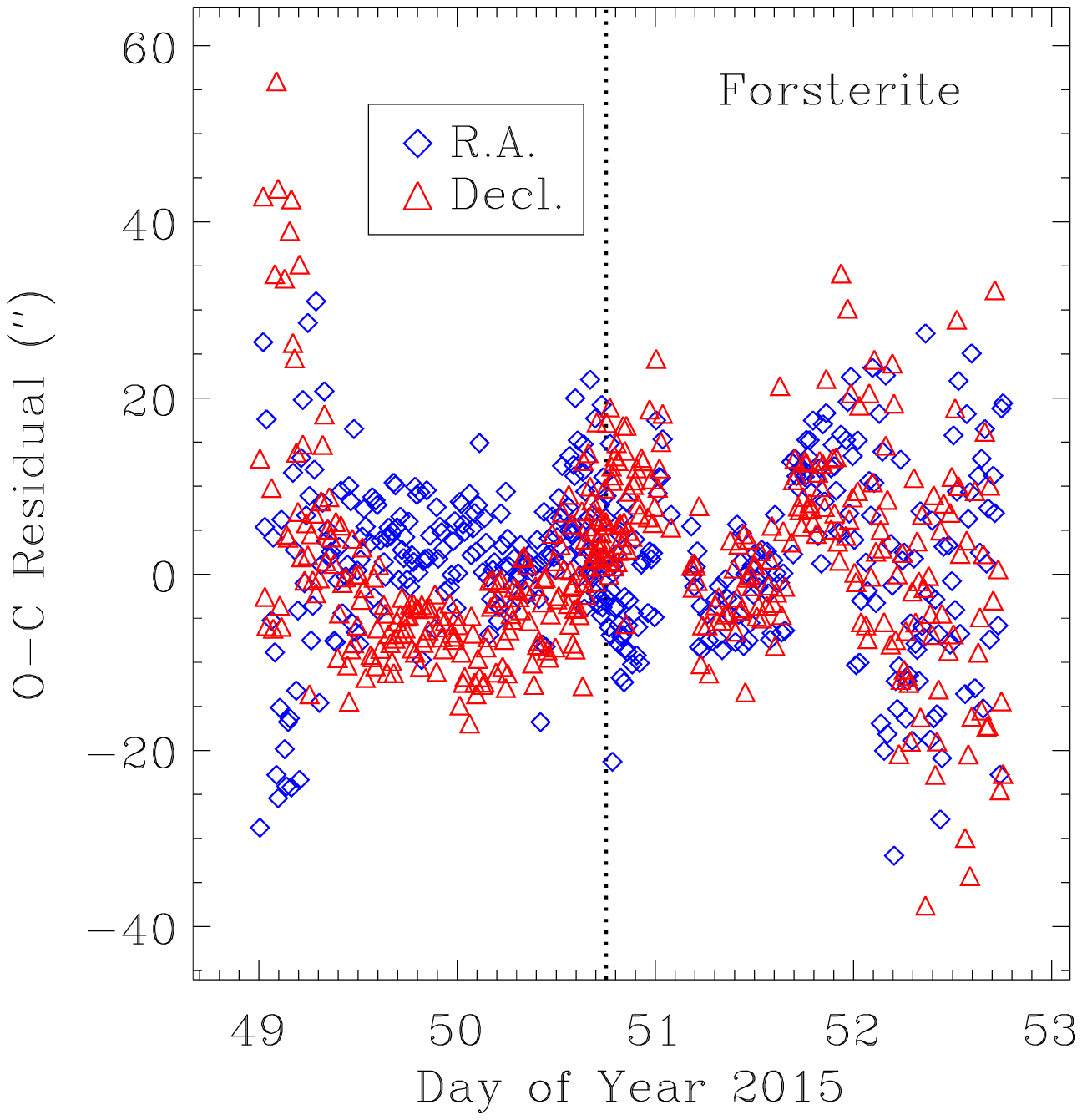} \\
    \centering\small ~~(c)
  \end{tabular}
  \caption{
Plots of O$-$C residuals in right ascension and declination as functions of time in different orbit determinations. The left panel (a) shows residuals from the pure gravitational solution, the middle one (b) shows residuals from the non-gravitational solution based on an isothermal water-ice sublimation model, and the right one (c) are residuals from the non-gravitational solution with a forsterite sublimation model. A sinusoidal shape in the left panel is clearly seen. Although significant residuals still exist, the solution with a water-ice sublimation model overall gives the best RMS and removes the peculiar trends presented in the left panel. Each panel marks the perihelion of 2015 D1 by a vertical dotted line. Note that the three panels have different ordinate scales.
  \label{fig:orb_resid}
  }
\end{figure}

 We investigate the color of 2015 D1 based upon filter magnitude differences. Clear magnitudes were determined at the time of orange/blue images by least squares interpolation between the nearest clear measurements. The clear magnitude errors were estimated from the neighboring clear magnitude errors and were combined with the orange/blue magnitude errors using standard error propagation techniques to give a total magnitude uncertainty on the color. Figure \ref{fig:color_2015D1} shows magnitude differences as functions of time and heliocentric distances. The non-zero clear $-$ blue and clear $-$ orange magnitude differences, particularly pre-perihelion, suggest that the color of 2015 D1 was distinctly different from the color of the Sun. However, the color was generally approaching to the solar color gradually as time evolved, despite some scatter around perihelion.
 
 We first examine whether the color can be attributed to thermal emission. We approximate dust grains as greybodies. Hence the effective temperature $T_{\mathrm{eff}}$ is given by
 \begin{equation}
 T_{\mathrm{eff}} = C_{\mathrm{S}} \left[ \left( 1 - A_{\mathrm{p}} \right) \frac{S_{\odot}}{4\epsilon \sigma r_{\mathrm{h}}^{2}} \right]^{\frac{1}{4}}
 \label{eq_Teff}
 \end{equation}
 \noindent in which $C_{\mathrm{S}}$ is superheat, $A_{\mathrm{p}} = 0.04$ is a nominal albedo for cometary nucleii (e.g., Lamy et al. 2004), and $S_{\odot} = 1361$ W m$^{-2}$ is the solar constant (Kopp \& Lean 2011), $\epsilon$ is the effective emissivity, assumed to be unity, and $\sigma = 5.6704\times10^{-8}$ W m$^{-2}$ K$^{-4}$ is the Stefan-Boltzmann constant. The influence from thermal radiation is evaluated by
\begin{equation}
\frac{F_{\mathrm{th}} }{F_\mathrm{sc} } = \frac{ r_{\mathrm{h}}^2 \int \mathfrak{T} B_\lambda \left(T_{\mathrm{eff}}\right) \mathrm{d} \lambda}{\int \mathfrak{T} A_{\mathrm{p}} \phi \left(\alpha\right) F_{\odot, \lambda} \mathrm{d} \lambda}  ,
\label{fth_to_fsc}
\end{equation}
\noindent where $F_{\mathrm{th}}$ is the thermal emission flux, $F_\mathrm{sc}$ is the flux due to scattering sunlight, $F_{\odot, \lambda}$ is the solar irradiance spectrum observed at 1 AU, $B_{\lambda}$ is the thermal emission flux from Planck's law, $\mathfrak{T}$ is the effective transmissivity of a given filtered optical system, and $\lambda$ is wavelength. We calculate $\mathfrak{T}$ for C2/C3 orange, C3 clear, and C3 blue filters based upon the information provided on the LASCO calibration page\footnote{\url{http://lasco-www.nrl.navy.mil/index.php?p=content/level_1/lascocal_index}}. The 1985 Wehrli solar spectrum\footnote{\url{http://rredc.nrel.gov/solar/spectra/am0/wehrli1985.new.html}} is used in this estimate. Unfortunately there are no available data which can constrain $C_{\mathrm{S}}$; it is a function entangled with $T_{\mathrm{eff}}$ and dust grain size (Gehrz \& Ney 1992). For simplicity, we assume $C_{\mathrm{S}} \equiv 1.2$, which is approximately the average of superheat values of all types of comets listed in Gehrz \& Ney (1992).

\begin{figure}
\begin{center}
\includegraphics[scale=0.9]{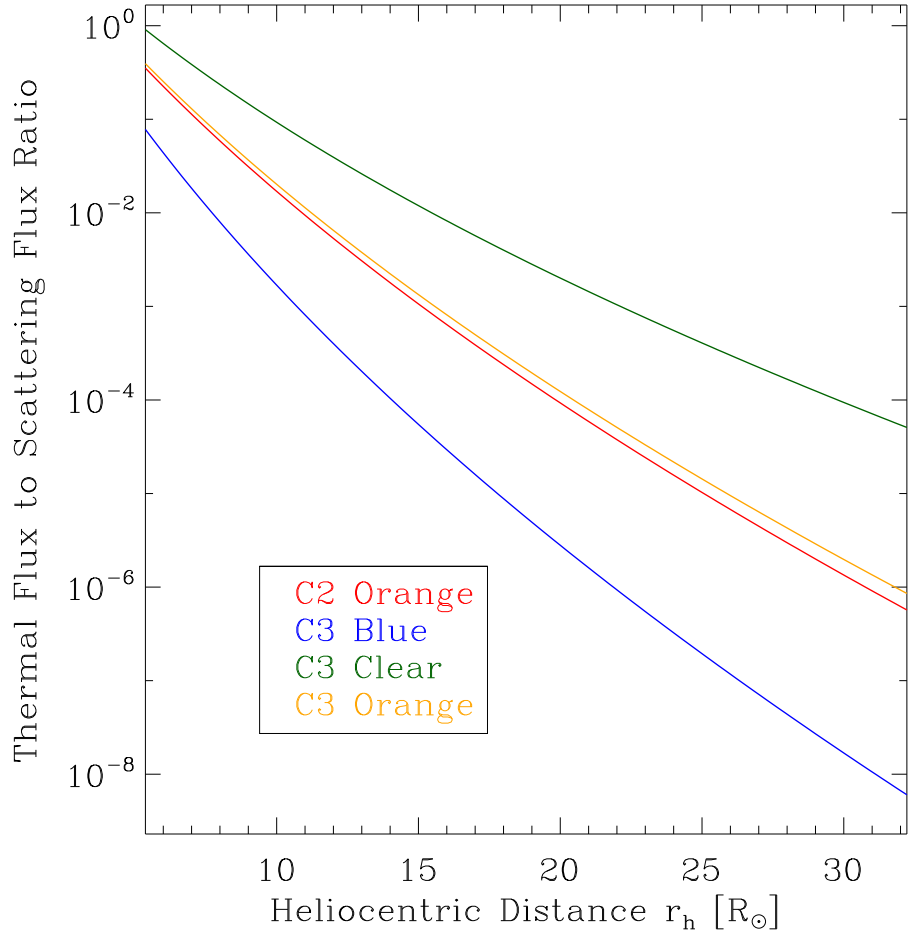}\\
\caption{
Assessment of influences from thermal radiation approached by examining the ratios of thermal emission flux to solar continuum flux $F_{\mathrm{th}}/F_{\mathrm{sc}}$ as a function of heliocentric distance $r_{\mathrm{h}}$, observed in different \textit{SOHO}/LASCO bandpasses. The closer to the Sun, the more influential thermal radiation is.
\label{fig:mod_therm}
}
\end{center}
\end{figure}

We compute $F_{\mathrm{th}} / F_{\mathrm{sc}}$ observed by LASCO blue, clear and orange filters. The simulation results are illustrated in Figure \ref{fig:mod_therm}, from which we can see that influence from thermal radiation emission was very limited for all LASCO filters at $r_{\mathrm{h}} \gtrsim 10~R_{\odot}$. Around perihelion, while C3 clear data would be affected by thermal radiation the most significantly, other filters would still receive negligible thermal radiation. However, since dust grains experience sublimation at small $r_{\mathrm{h}}$, the actual equilibrium temperature would therefore be lower than what Equation (\ref{eq_Teff}) gives (c.f. Kimura et al. 2002). Hence, we think that the effect due to thermal radiation emission can be ignored.

Intuitively, were the color of 2015 D1 due to thermal radiation entirely, it is expected that the comet would appear redder at smaller $r_{\mathrm{h}}$, which is not observed. Furthermore, as indicated in Figure \ref{fig:mod_therm}, the comet should always appear the brightest in C3 clear images, mediocre in C2/C3 orange, and the faintest in C3 blue data, obviously contradictory to the LASCO observation (Figure \ref{fig:color_2015D1}). We can therefore conclude that thermal radiation emission is not responsible for the observed color of the comet. 

\begin{figure}
  \centering
  \begin{tabular}[b]{@{}p{0.45\textwidth}@{}}
    \centering\includegraphics[scale=0.6]{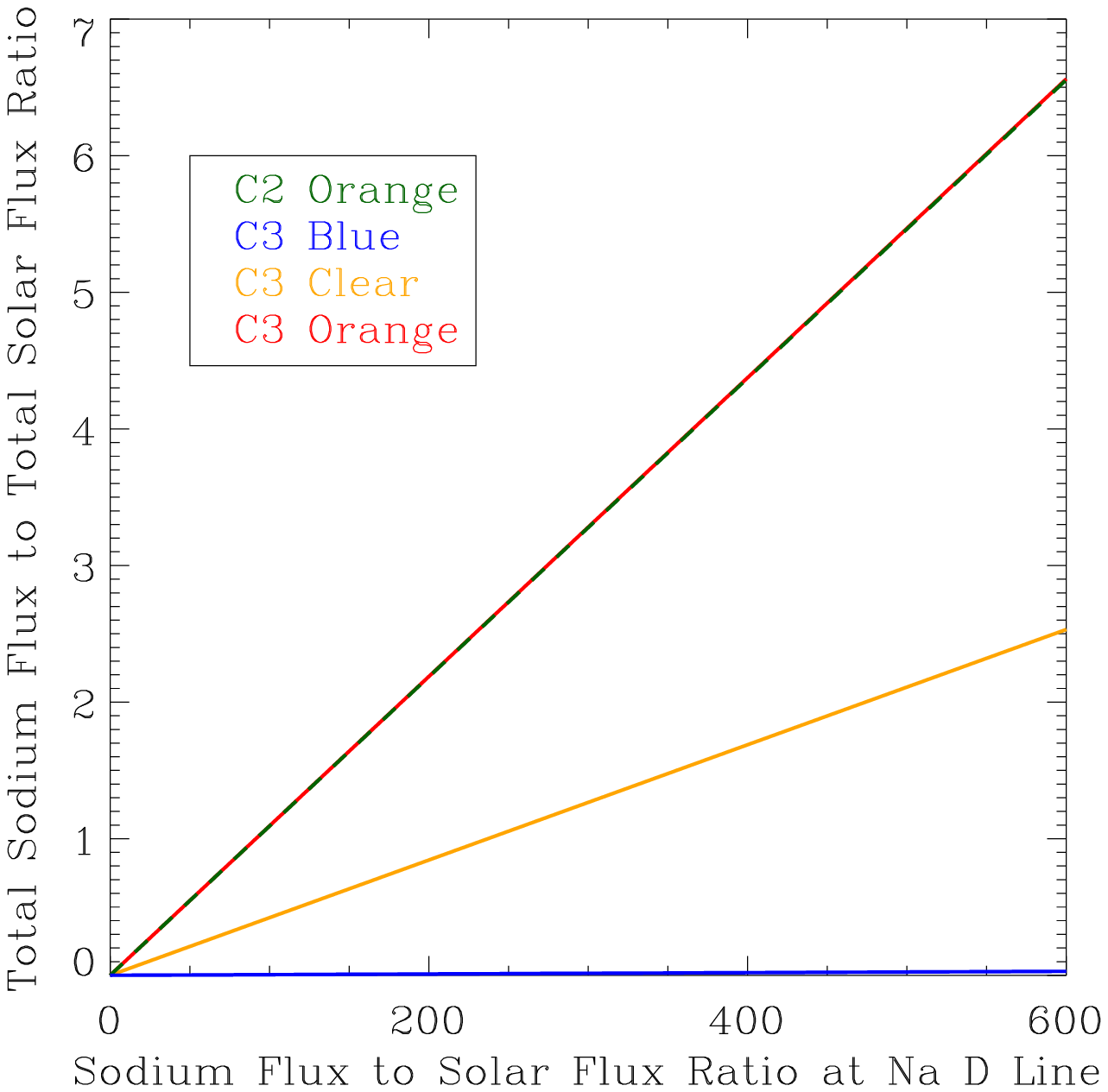} \\
    \centering\small (a)
  \end{tabular}%
  \quad
  \begin{tabular}[b]{@{}p{0.45\textwidth}@{}}
    \centering\includegraphics[scale=0.6]{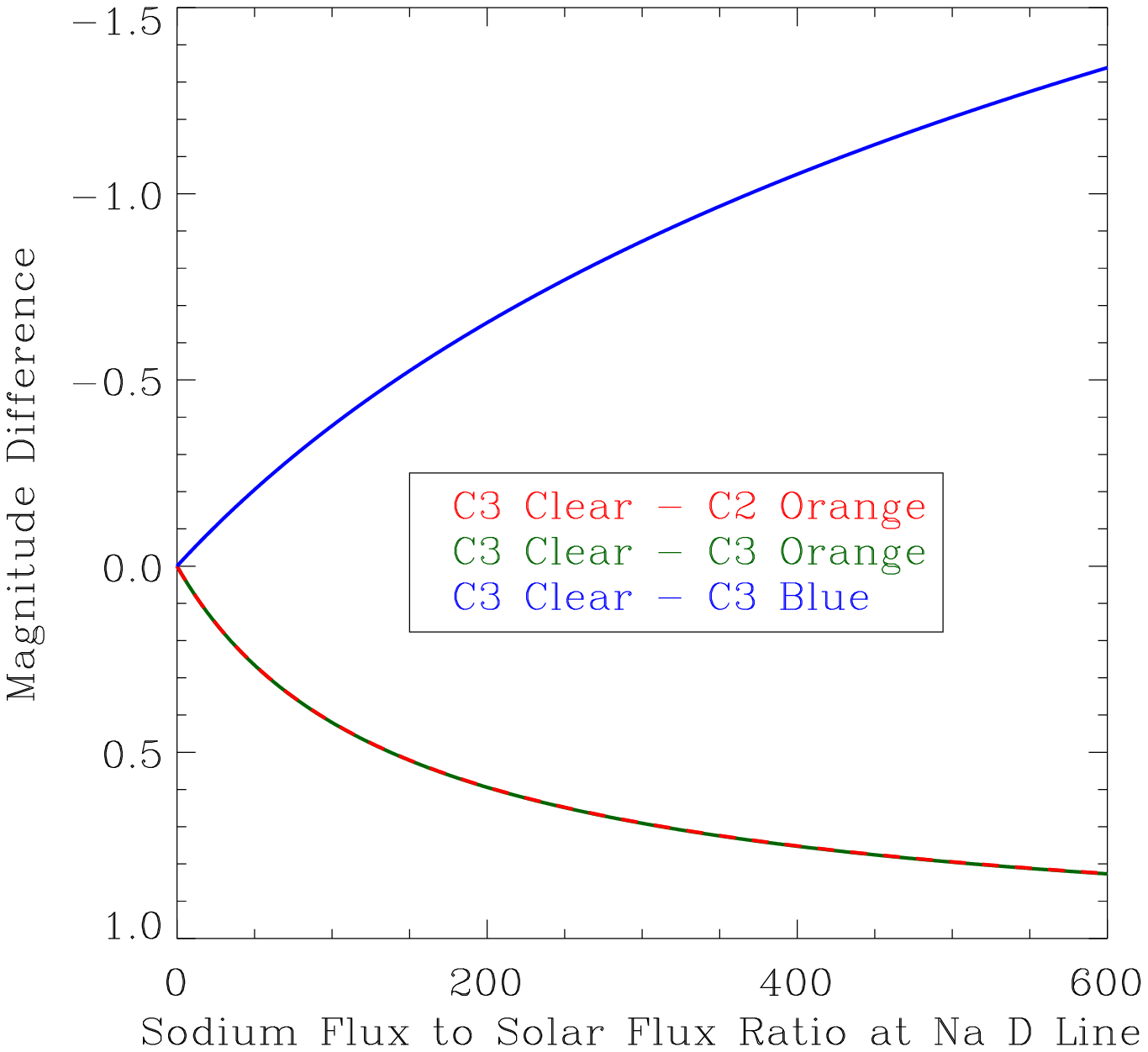} \\
    \centering\small (b)
  \end{tabular}
  \caption{
  Modeled influences from sodium emission observed in different \textit{SOHO}/LASCO bandpasses. Details are discussed in Section \ref{disc_color}. Note that C2 orange and C3 orange show no obvious differences and therefore overlap each other.
  \label{fig:mod_Na}
  }
\end{figure}

 We next investigate how sodium emission will influence the color of 2015 D1, since this effect is prominent when a comet nears the Sun. Similar to the method described in Knight et al. (2010), we add a synthetic rectangular sodium flux $F_{\mathrm{Na}}$ with varying intensity to the solar spectrum centered at Na D-line $\lambda = 5985$ \AA~with a fixed width $\Delta \lambda_{\mathrm{Na}} = 10$ \AA. Since $\Delta \lambda_{\mathrm{Na}}$ is very small, $\overline{\mathfrak{T}}_{\mathrm{Na}}$, the mean effective transmission around the Na D-line within $\Delta \lambda_{\mathrm{Na}}$, can be utilized for simplification, such that the modeling magnitude difference between filter $i$ and $j$ now becomes
 \begin{equation}
 \Delta m_{i,j} = -2.5 \log \left[ \left( \frac{ F_{\mathrm{Na}} \overline{\mathfrak{T}}_{\mathrm{Na}, i} \Delta \lambda_{\mathrm{Na}} + \int \mathfrak{T}_{i}  F_{\odot, \lambda}  \mathrm{d}\lambda } { F_{\mathrm{Na}} \overline{\mathfrak{T}}_{\mathrm{Na}, j} \Delta \lambda_{\mathrm{Na}} + \int \mathfrak{T}_{j} F_{\odot, \lambda}  \mathrm{d}\lambda} \right) \left( \frac{  \int \mathfrak{T}_{j} F_{\odot, \lambda} \mathrm{d}\lambda} {  \int \mathfrak{T}_{i} F_{\odot, \lambda} \mathrm{d}\lambda} \right)  \right] ,
 \label{eq_model_dmag}
 \end{equation}
 \noindent whereby we obtain the results shown in Figure \ref{fig:mod_Na}. Although only the flux due to the solar continuum and the flux due to the sodium emission are taken into consideration, the modeled magnitude differences can be matched by varying the intensity of the sodium emission. For instance, an intensity of sodium emission $\sim$100 times stronger than the solar continuum at 5985 \AA~corresponds to an apparent magnitude $\sim$0.4 mag brighter in the C2/C3 orange filters and $\sim$0.3 mag fainter in the C3 blue relative to the C3 clear filter, which was exactly the color of 2015 D1 around UT 2015 February 19.9 (DOY = 50.9). We thus think that the sodium emission was a plausible mechanism to account for the color of the comet observed in LASCO cameras.
 
The magnitude difference between the clear and orange filters decreased as the comet approached perihelion, indicating depletion of sodium emission and the coma becoming increasingly more dusty, i.e. $\delta_{90}$ increasing. Since the corrections from $\delta_{90}$ are generally comparable to the uncertainties in the magnitude data, it is not meaningful to apply a temporally varying $\delta_{90}\left( t \right)$ to correct for the phase function. Thus we take the mean magnitude differences to derive $\left \langle \delta_{90} \right \rangle$ for clear, orange and blue filters with Figure \ref{fig:mod_Na}b respectively. For C3 blue filter we have $\delta_{90} = 88$, however, several typical cometary emission lines, e.g., $\mathrm{C}_{2}$, CN, etc., would be transmittable through the bandpass and likely lower this value considerably. Thus a conservative $\delta_{90} = 10$ is used.

\subsection{Ejection of Dust Grains}
\label{sim_morph}

To understand the morphology of the comet as well as the properties of the remaining debris cloud, we employed a Monte Carlo dust model similar to the one used in Ye \& Hui (2014) to generate synthetic images of the comet. During initial tests we noted the unique challenges for the case of 2015 D1. Firstly, the low spatial resolution of LASCO C3 images prevents us from obtaining information about the surface brightness profile of the cometary tail. Particularly, the pre-perihelion tail stretched $\le 4$ pixels (i.e. $\lesssim 4\arcmin$) in these images, too small for model comparison. By the time the comet appeared in LASCO C2, the tail had begun dimming already. Secondly, for ground-based observations, the combined effect from the nebulous nature of the remnant, the lack of a central condensation as a reference point, and the large uncertainty of the comet's position ($\sim$1\arcmin~or $\sim$100 pixels in images from Xingming) make it very difficult to directly assess the goodness of the model. Therefore, we only focus on matching the general shape of the tail/remnant starting from around perihelion. Nevertheless, thanks to the small heliocentric distances at which the dust grains were released, the often-significant divergences between different sets of parameters made it relatively easy to identify implausible solutions.

The dynamics of cometary dust grains are determined by the $\beta$ parameter, the ratio between the solar radiation force and the gravitational force exerted by the Sun, and the initial ejection velocity. The ratio $\beta$, dust grain radius $a$ and bulk density of dust $\rho_{\mathrm{d}}$ are related by
\begin{equation}
\beta = \frac{\mathcal{C}}{\rho_{\mathrm{d}} a}
\label{eq_beta},
\end{equation}
\noindent where $\mathcal{C} = 5.95 \times 10^{-4}$ kg m$^{-2}$ is a proportionality constant. After trials with various parameter valuations, we found that the post-perihelion shape of the tail/remnant was predominantly controlled by the generation of small dust particles. Hence, in the following, we use the dust ejection model by Crifo \& Rodionov (1997) and the upper limit of dust size $a_{\max} \sim 1$ cm. Assuming a typical bulk density $\rho_{\mathrm{d}} = 0.4$ g cm$^{-3}$ (e.g., Richardson et al. 2007), we have $\beta_{\min} \sim 1.5\times10^{-4}$.

An immediate question is the duration of dust ejection: did the nucleus split instantaneously (such that the dust ejection ceased shortly after the comet's flareup), or did the disintegration process last for some period of time? We thus consider three scenarios:

\begin{enumerate}
 \item All dust grains were impulsively ejected at the start of the flareup at $\Delta t \sim -1$ hr (impulsive ejection);
 \item The dust grains were ejected from the start of the flareup to the peak brightness, i.e. $-1 \lesssim \Delta t \lesssim +3$ hrs (short semi-impulsive ejection); and
 \item The dust grains were ejected from the start of the flareup to the time when signs of nucleus disintegration were seen in \textit{SOHO} images, i.e. $-1 \lesssim \Delta t \lesssim +1$ day (long semi-impulsive ejection).
\end{enumerate}

The simulated particles, isotropically released, were generated using both sets of the orbital elements in Table \ref{orbit} during initial tests. A modified MERCURY6 package (Chambers 1999) was used to integrate all particles to observation epochs using the Bulirsch-Stoer integrator (Bulirsch 1972, Stoer 1972). Radiation forces are included in the code. Also included are gravitational perturbations from the eight major planets, although these cast no visible influence on modeling 2015 D1. We then calculated the positions of simulated particles with respect to \textit{SOHO} or the Earth at epochs of interest to produce the shapes of the dust ensembles.

\begin{figure}
\begin{center}
\includegraphics[scale=2.0]{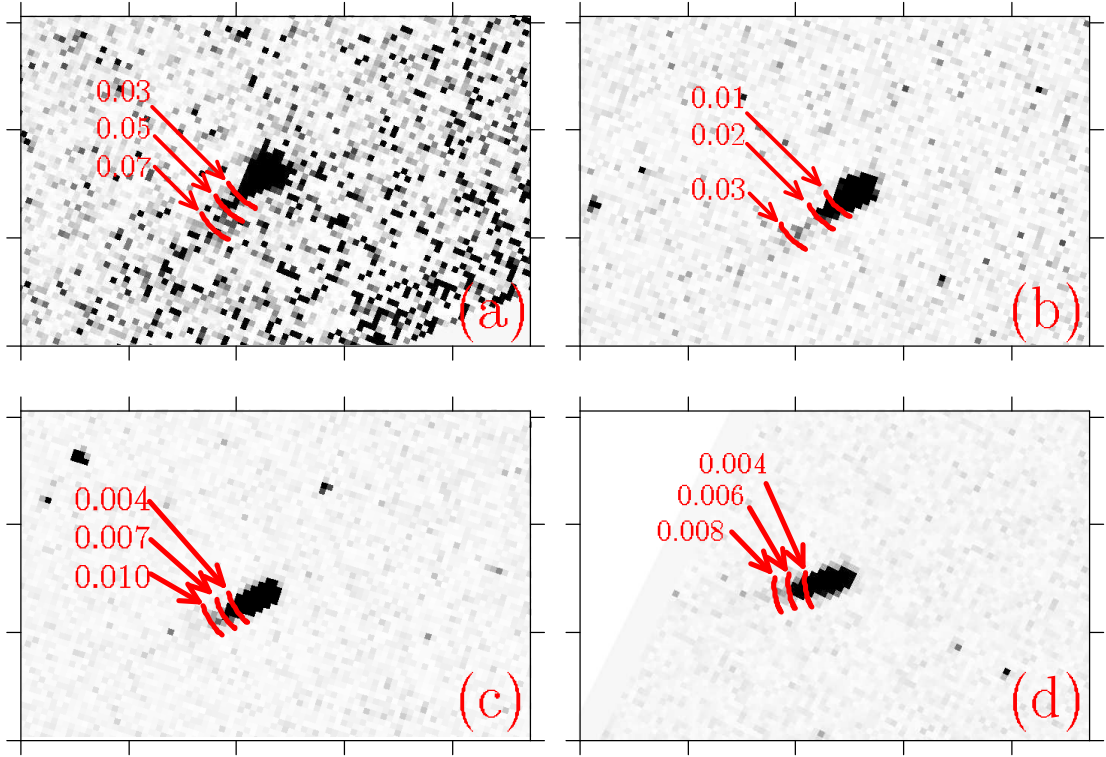}
\caption{Termini of dust grains at different $\beta_{\max}$ in four selected \textit{SOHO} images: (a) UT 2015 February 20 07:18; (b) February 20 18:06; (c) February 21 5:42; and (d) February 21 15:18. The dust models shown here were generated using the impulsive ejection model. The difference between impulsive ejection and short/long semi-impulsive ejection is not distinguishable in \textit{SOHO} images. Ticks are plotted in the interval of $10$\arcmin, and $\beta$ values are indicated on the plots. The images are oriented such that north is up and east is left.}
\label{fig:mdl-soho}
\end{center}
\end{figure}

\begin{figure}
\begin{center}
\includegraphics[scale=0.9]{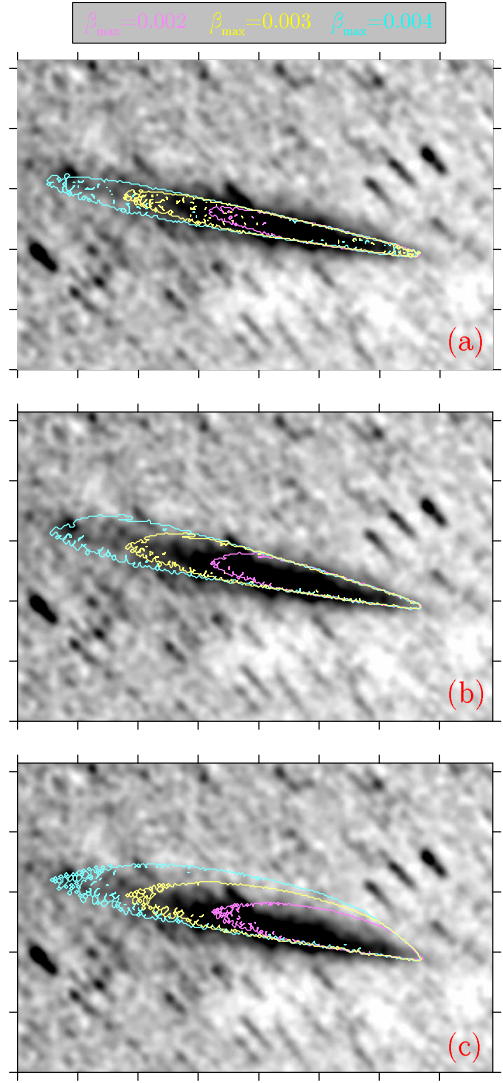}
\caption{Xingming image on 2015 March 4 overlaid with the best dust models (contours) of (a) impulsive ejection ($t - t_\mathrm{P} \sim -1$ hr); (b) short semi-impulsive ejection ($-1 \lesssim t - t_\mathrm{P} \lesssim +3$ hrs); and (c) long semi-impulsive ejection ($-1 \lesssim t - t_\mathrm{P} \lesssim +1$ day). The results for March 8 and 15 are largely identical. Dust models are translated $\sim$3\arcmin~northwest to counter the offset presumably introduced by an imperfect ephemeris. The model agrees with the observation for the cases of impulsive and short semi-impulsive ejections (i.e. ejection duration $<0.1$~day). Ticks are plotted in the interval of $10$\arcmin. The images are oriented so that north is up and east is left.}
\label{fig:mdl-20150304}
\end{center}
\end{figure}

 During tests, by visual inspection, no distinction between the modeled shapes from different sets of orbital solutions was detected. We think that different orbital solutions affect little the morphological analysis, and therefore applied the solution without $\mathcal{A}_{1}$. We tested $\beta_{\max}$ from $5 \times 10^{-4}$ to 0.5 using a logarithmically varying interval (i.e., steps of 10$^{-4}$ for $\beta_{\max} \sim 10^{-4}$, steps of 10$^{-3}$ for $\beta_{\max} \sim 10^{-3}$, etc.). We selected eight \textit{SOHO} images from UT 2015 February 19 20:06 to February 21 15:18, each separated by about 6 hrs (except the first two images, from UT February 19 20:06 and February 20 07:18 respectively, are separated by 11 hrs, as the comet was obstructed by the pylon of the coronagraph), and Xingming images from March 4, 8 and 15 (observations from March 9 were dismissed due to a bright background star) for model matching. For \textit{SOHO} data, the synthetic images are essentially a set of segments due to the low resolution of \textit{SOHO}. The goodness of the model is therefore assessed by comparing the distance traveled by different sizes of dust to the observed length of the tail (Figure~\ref{fig:mdl-soho}). For Xingming data (Figure \ref{fig:mdl-20150304}), the shape of the remnant seems reproducible; however, we notice that the modeled debris cloud is constantly $\sim$3\arcmin~southeast of the actual observed cloud, a phenomenon we attribute to imperfect orbit determination. Note that this was present no matter which orbital solution (our own, JPL, MPC) was used. The positions of the simulated particles were therefore translated $\sim$3\arcmin~in the northwest direction to align the synthetic images to the observations (Figure~\ref{fig:mdl-20150304}). 

\begin{figure}
\begin{center}
\includegraphics[scale=1.5]{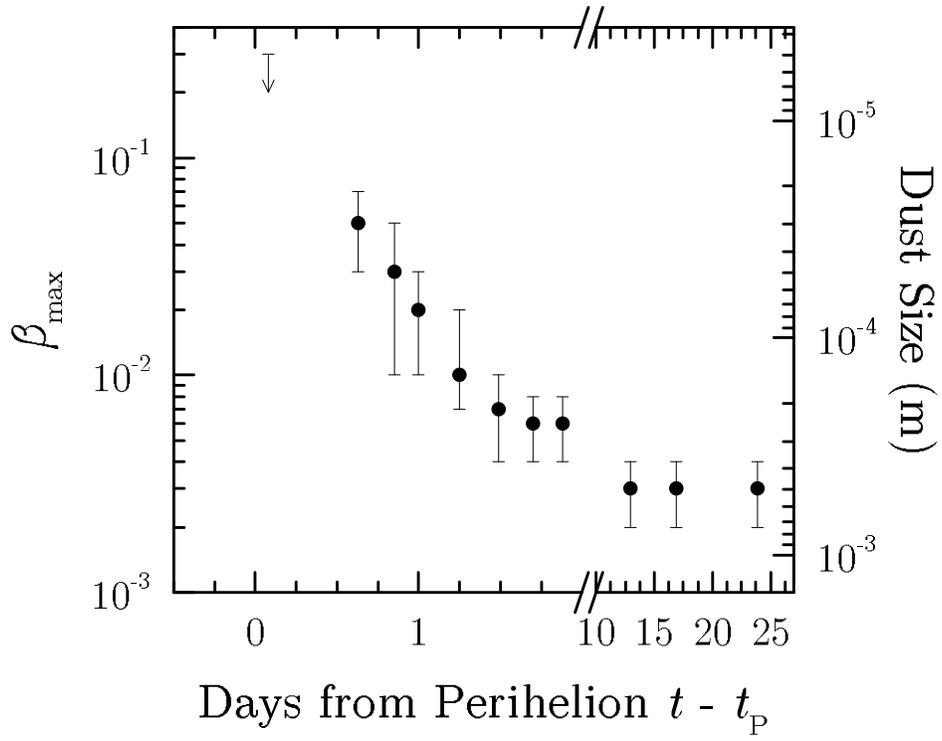}
\caption{The temporal decrease of $\beta_{\max}$ as seen in \textit{SOHO} and Xingming data.}
\label{fig:mdl-beta}
\end{center}
\end{figure}

The ejection duration is constrained by Xingming data, indicating a quasi-impulsive ejection of the dust within 0.1 day (see Figure~\ref{fig:mdl-20150304}) and endorsing the idea that the destruction occurred immediately after the flareup. This is consistent with the analysis by Sekanina (2015). The \textit{SOHO} images, suffering from low spatial resolutions, failed to allow a clear separation of different ejection durations, although the length of the tail provides a reliable constraint to the lower size limit of the optical dust. An increasing trend of $a_{\min}$ is clearly noticeable (Figure~\ref{fig:mdl-beta}). The freshest dust grains had $a_{\min} \sim 10$ $\mu$m; it increased at a rate of $\dot{a}_{\min} \sim 10^{-1}$ mm day$^{-1}$, and stabilized at $\sim$0.5 mm. This may be explained by observational bias: the smallest dust grains in the debris cloud quickly dispersed and thus dimmed beyond the observation threshold, and the debris cloud was expanding due to solar radiation pressure without replenishment of dust particles. Conversely, larger dust grains expanded more slowly and remained observable for a longer period of time.

 \subsection{Size Estimate}
 
 \subsubsection{Nucleus Size from Photometry}
 \label{disc_sz_phot}
\begin{figure}
\begin{center}
\includegraphics[scale=0.9]{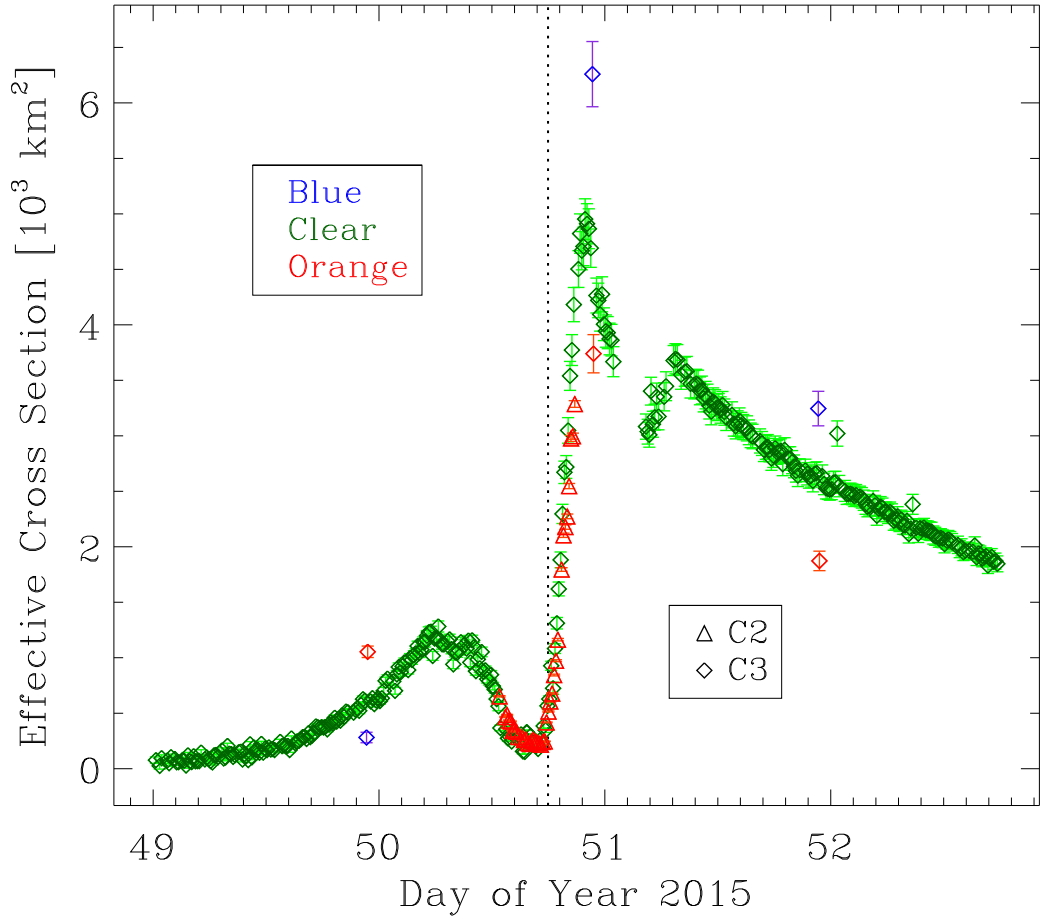}\\
\caption{
Temporal variation of effective cross-section area of 2015 D1 against time from the LASCO observation. The vertical dotted line labels the perihelion moment. Point symbols correspond to telescopes and points are color coded according to filters.
\label{fig:xs_2015D1}
}
\end{center}
\end{figure}

 We can estimate the nucleus size of 2015 D1 from the \textit{SOHO} photometric data. The heliocentric magnitude of the comet due to the dust grains which reflect sunlight can be extracted, similar to Equation (\ref{eq_heliomag_gas}), by the formula
 \begin{equation}
 H_{V,\mathrm{dust}} = H_{V} + 2.5 \log \left(1 + \frac{1}{\delta_{90}} \right) .
 \end{equation}
 \noindent Then the effective cross-section $C_{\mathrm{e}}$ of the comet can be calculated as
 \begin{equation}
 C_{\mathrm{e}} = \frac{\pi r_{\mathrm{h}}^{2} }{A_{\mathrm{p}}} 10^{-0.4 \left(H_{V,\mathrm{dust}} - m_{\odot, V} \right)}.
 \label{eq_xs}
 \end{equation}
 \noindent Here, $m_{\odot, V} = -26.74$ is the apparent $V$ band magnitude of the Sun. We still use $A_{\mathrm{p}} = 0.04$ for the dust grains. Figure \ref{fig:xs_2015D1} shows $C_{\mathrm{e}}$ as a function of time. We assume that the optically thin coma is made of spherical dust grains whose radii range from $a_{\min}$ to $a_{\max}$, and that they obey a power-law size distribution, $\mathrm{d}N \propto a^{-\gamma} \mathrm{d}a$, with a constant $\gamma$. Then the effective nucleus radius $R_{\mathrm{N}}$ can be solved by
 
 \begin{equation}
 R_{\mathrm{N}} = \left[ \frac{1}{\pi} \left(\frac{3 - \gamma}{4 - \gamma} \right) \left(\frac{a_{\max}^{4-\gamma} - a_{\min}^{4-\gamma}}{a_{\max}^{3-\gamma} - a_{\min}^{3-\gamma}} \right) C_{\mathrm{e}} \right]^{1/3} .
 \label{eq_RN}
 \end{equation}
 
 From the morphological analysis in Section \ref{sim_morph} we have $10~\mu\mathrm{m} \lesssim a \lesssim 1$ cm around perihelion. We can constrain $\gamma$ from the uniform decline in $C_{\mathrm{e}}$ starting from $\Delta t \sim 0.6$ day (DOY $\sim$ 51.3) until the end of the LASCO observation by assuming that the decline was completely attributed to faster dispersions of smaller dust grains accelerated by solar radiation forces. The relationship between $C_{\mathrm{e}}$ and the dust size distribution is
 \begin{equation}
 C_{\mathrm{e}} \left( t \right) = \mathfrak{C} \left[ a_{\min}^{3 - \gamma} \left( t \right) - a_{\max}^{3 - \gamma} \right]
 \label{eq_Ce},
 \end{equation}
 \noindent where $\mathfrak{C}$ is an unknown constant that does not affect the calculation. We know $a_{\min} \left( t \right)$ in the same interval of time from Figure \ref{fig:mdl-beta}. A best fit to Equation (\ref{eq_Ce}) by MPFIT (Markwardt 2009) yields $\gamma = 3.16$. It is relatively insensitive to $a_{\max}$ and $C_{\mathrm{e}}$. For instance, changing $a_{\max}$ from 5 mm to $10^{2}$ m varies $\gamma$ from 3.10 to 3.30. We are confident that distributions with $\gamma = 3.2 \pm 0.1$ encompass the likely range of parameter uncertainties. In comparison, distributions with $3.5 \leq \gamma \leq 4.1$ have been found for a large number of comets (e.g., Sitko et al. 2011), but $\gamma = 3.2 \pm 0.1$ is not uncommon (e.g., Fulle 2004). 
 
 Around perihelion, Equation (\ref{eq_RN}) yields $R_{\mathrm{N}} \simeq 0.11\pm0.01$ km. Taking into account different assumptions about the albedo (e.g., Lamy et al. 2004 estimates an uncertainty of $\pm$0.017) yields $R_{\mathrm{N}} \simeq 0.11_{-0.02}^{+0.04}$ km, which likely encompasses the original nucleus size.

 \subsubsection{Nucleus Size from Non-Gravitational Effect}
 \label{disc_sz_ng}
  
  Whipple (1950) shows that the nucleus mass $M_{\mathrm{N}}$ can be inferred from the non-gravitational acceleration as a result of momentum conservation. The composite non-gravitational parameter, $\mathcal{A} = \sqrt{\mathcal{A}_{1}^{2} + \mathcal{A}_{2}^{2} + \mathcal{A}_{3}^{2}}$, is connected to the non-gravitational acceleration by $\mathfrak{A} \left(r_{\mathrm{h}} \right) =  \mathcal{A} g\left( r_{\mathrm{h}} \right)$, where $g\left(r_{\mathrm{h}} \right)$ is the dimensionless empirical non-gravitational momentum transfer law from an isothermal water-ice sublimation model (Marsden et al. 1973), which we exploited in determining $\mathcal{A}_{1}$. However, the actual mechanism the nucleus of 2015 D1 suffered might well be too complicated to be described by any simple models. Additionally, evidence suggests that sublimation of olivines (e.g., forsterite) has taken place around perihelion. But the contribution to the non-gravitational effect is believed to be very limited (see Section \ref{disc_mloss}). Given the high uncertainties in the astrometric data, we still apply the empirical law $g\left(r_{\mathrm{h}} \right)$ from the water-ice sublimation model.

  We thus have
 \begin{equation}
 M_{\mathrm{N}} = \kappa \frac{Q \mu v}{\mathfrak{A}} , 
 \label{eq_mass_nuc}
 \end{equation}
 
 \noindent where $Q\left( r_{\mathrm{h}} \right)$ is the production rate of the dominant mass loss material, i.e. water-ice, having molecular mass $\mu\left(\mathrm{H_{2}O} \right) = 18$ u, and $\kappa$ is the dimensionless collimation efficiency, with $\kappa = 0$ for isotropic emission and $\kappa=1$ for perfect-collimated ejection, and $v$ is the outflow speed of gas as a function $r_{\mathrm{h}}$, which is ill-defined when $r_{\mathrm{h}}$ is small. Applying relationships such as those given by Delsemme (1982) and Biver et al. (1998) to a near-Sun scenario is probably inappropriate. Instead, we approximate $v$ as thermal speed
 \begin{equation}
 v_{\mathrm{th}}\left(r_{\mathrm{h}} \right) = \left[\left( \frac{3 k_{\mathrm{B}}}{\mu} \right)^4 \frac{\left(1-A_{\mathrm{p}} \right)S_{\odot}}{4\epsilon \sigma r_{\mathrm{h}}^{2}} \right]^{1/8} ,
 \label{eq_outgass_speed}
 \end{equation}
 \noindent where $k_{\mathrm{B}} = 1.3806\times10^{-23}$ J K$^{-1}$ is the Boltzmann constant. Assuming $A_{\mathrm{p}} = 0.04$ and $\epsilon = 1$, for water-ice sublimation, Equation (\ref{eq_outgass_speed}) is simplified to $v_{\mathrm{th}} \left(r_{\mathrm{h}} \right) = 0.62~r_{\mathrm{h}}^{-0.25}$ km s$^{-1}$, where $r_{\mathrm{h}}$ is expressed in AU.
 
  By no means can $\kappa$ be constrained from the observations, and we somewhat arbitrarily adopt $\kappa = 0.5$. There is no constraint on the gas production rate of 2015 D1 either, therefore the empirical law for long-period comets by Sosa \& Fern\'andez (2011) is applied. We use the magnitude only due to gas emission calculated by Equation (\ref{eq_heliomag_gas}) with $\mathcal{G} \equiv 1$. Hence Equation (\ref{eq_mass_nuc}) yields a mean nucleus mass $\left \langle M_{\mathrm{N}} \right \rangle \simeq \left( 5.1 \pm 3.3 \right) \times10^{8}$ kg, much smaller than the masses of long-period comets studied by Sosa \& Fern\'andez (2011), by 4 orders of magnitude. Assuming $\rho_{\mathrm{d}} = 0.4$ g cm$^{-3}$, we have its effective nucleus radius $R_{\mathrm{N}} = \sqrt[3]{3 M_{\mathrm{N}} / \left( 4\pi\rho_{\mathrm{d}} \right)} \simeq 67 \pm 15$ m. 
  
  However, in Section \ref{disc_sz_phot} the nucleus size is estimated to be $0.11_{-0.02}^{+0.04}$ km in radius. Given the same $\rho_{\mathrm{d}}$, this yields $M_{\mathrm{N}} \simeq \left(1.3\mbox{--}5.5\right) \times 10^{9}$ kg, an order of magnitude larger than the mass derived from the non-gravitational effect. We consider the following reasons.
  
  \begin{enumerate}
  \item The estimated size from photometry includes all the constituents within the aperture, which occupies a spatial sphere of $\sim$$1.2\times10^{5}$ km in radius around 2015 D1's optocenter around perihelion, not the nucleus alone. Larger dust grains released earlier would stay in the aperture much longer than the entire passage in LASCO C3's FOV. Therefore photometric data give a decent estimate about the initial size, whereas the size estimate from the non-gravitational effect tends to give the size around perihelion. We thus expect the size estimated from photometry to be significantly larger than the one from the non-gravitational effect.
  \item The non-gravitational effect might come from a different mechanism other than from the isothermal water-ice sublimation model. As mentioned earlier, $\mathcal{A}_{1}$ in the orbital solution fails to remove an obvious leap in the astrometric residuals in declination around perihelion (Figure \ref{fig:orb_resid}). Together with the photometric data and morphological analysis, this broadly agrees that something catastrophic happened to the comet around perihelion. So one might expect that the non-gravitational acceleration emerged predominantly around perihelion, and a smooth and continuous model might have deviated from the fact.
  \item The empirical law of gas production rate by Sosa \& Fern\'andez (2011) has never been examined at small $r_{\mathrm{h}}$ and therefore it may be inappropriate to apply to 2015 D1 directly. Alternatively, 2015 D1's actual production rate might deviate from the empirical law, even though it might still hold at small $r_{\mathrm{h}}$ for other near-Sun comets.
  \end{enumerate}
  
  Given the substantial uncertainties associated with the behaviors of near-Sun comets, we think that both methods give acceptably consistent size estimates. We are confident that the nucleus mass of 2015 D1 was $\sim$10$^{8}$--10$^{9}$ kg before disintegration, e.g., much smaller than most comets studied by Earth-based observers near $r_{\mathrm{h}} \sim 1$ AU. Note that we only used the clear images because they had less potential sodium contamination than the orange images, and were acquired far more frequently than the blue images.

  \subsubsection{Constraints on Post-Perihelion Remnant}
  
 Using the more restrictive Lowell non-detection from the slightly smaller DCT FOV, we can also estimate the upper limit of an inactive outbound nucleus of 2015 D1 as follows. Assuming the comet has solar color, the SDSS r magnitude converts to V magnitude by $m_{r} = m_{V} – 0.16$ (Smith et al. 2002), yielding an upper limit of $m_{V} < 20.16$. We then derive its effective cross-section with Equation (\ref{eq_xs}) and use $R_{\mathrm{N}} = \sqrt{C_{\mathrm{e}}/\pi}$ to determine an upper limit to the remaining nucleus size as $R_{\mathrm{N}} \lesssim 0.6$ km. It is necessary to point out that we here apply the IAU H-G photometric system phase function by Bowell et al. (1989) for a bare nucleus to $\alpha = 0\degr$ with the slope parameter G = 0.15. 
 
 A radius of $R_{\mathrm{N}} \lesssim 0.6$ km is not particularly restrictive, considering that we have previously shown from both photometry and non-gravitational forces that the pre-perihelion nucleus size was $R_{\mathrm{N}} \lesssim 0.1$ km. Different assumptions about the albedo, the phase correction (e.g., Lagerkvist \& Magnusson 1990 shows that the slope parameter can be off from G = 0.15 by $\sim$$\pm$0.1), or the limiting magnitude (we estimated the comet could have been $\sim$1 mag fainter than the faintest stars due to trailing), still result in $R_{\mathrm{N}} \lesssim 0.28$ km in the most restrictive case.

We next consider the upper limit on an active nucleus radius during post-perihelion observations. First we estimate the upper limit to water production rate $Q$ based on our limiting magnitude following the empirical correlation found by Sosa \& Fern\'andez (2011), whereby we have $Q < 3.3\times10^{25}$ molecules s$^{-1}$. We next estimate the surface area necessary to produce this production rate at $r_{\mathrm{h}} = 0.596$ AU using the methodology of Cowan \& A’Hearn (1979) and translate this into an effective radius assuming the comet is active over a surface area corresponding to the effective cross section. This yields $R_{\mathrm{N}} \simeq 24$ m or 50 m for the subsolar or isothermal cases, respectively. Given the significant assumptions that go into this estimate, it is probable that any remaining active nucleus was less than 100 m in radius.

 \subsection{Mass Loss}
 \label{disc_mloss}
 
\begin{figure}
\begin{center}
\includegraphics[scale=1]{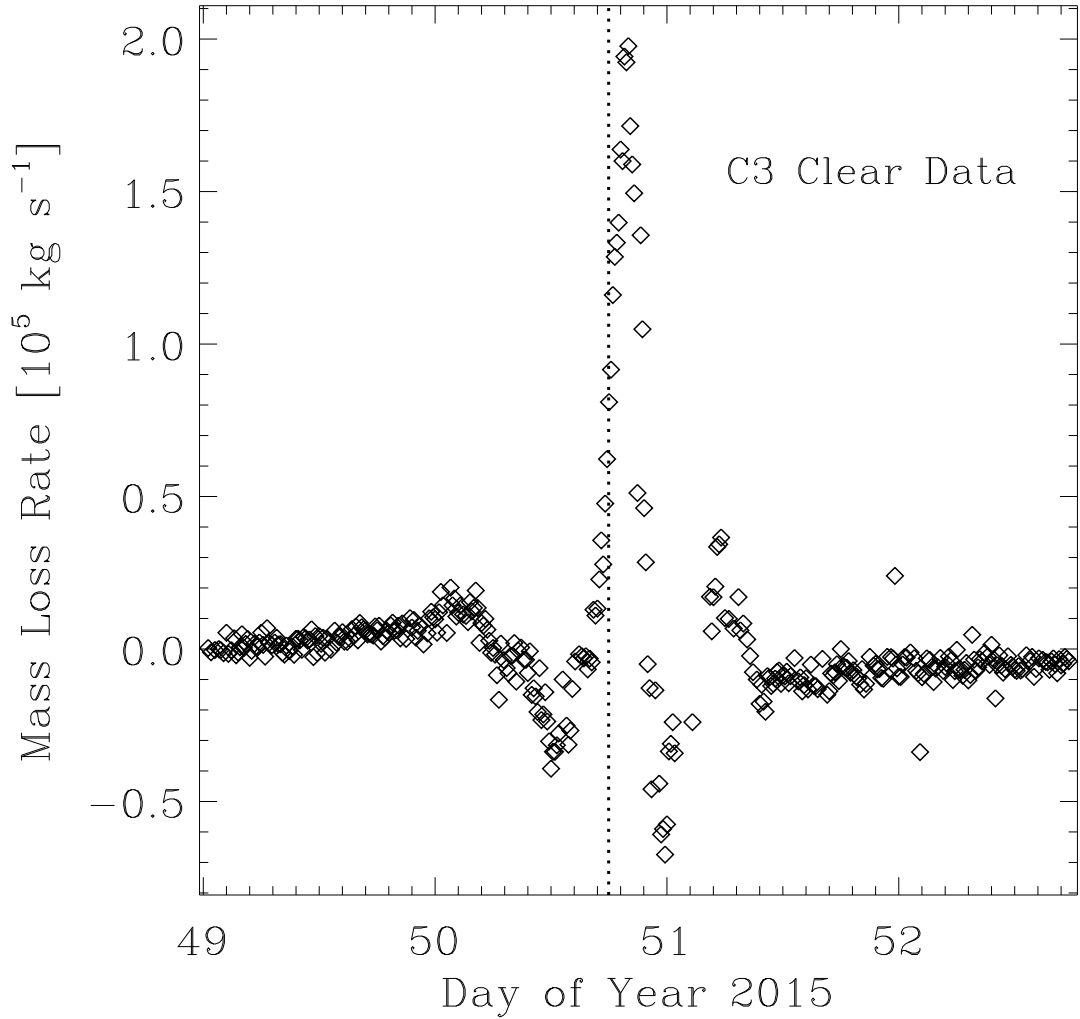}\\
\caption{
Mass loss rate calculated from photometric data. Only C3 clear data are used because of the adequate number. The perihelion moment is marked by a vertical dotted line in the middle of the graphic. Negative values in the plot should not be regarded as the authentic mass loss rate of the nucleus, but that the mass loss rate decreased due to particles drifting out of the photometric aperture.
\label{fig:mloss_rate}
}
\end{center}
\end{figure}

 We can investigate the mass loss of 2015 D1 from either photometry or the non-gravitational effect. Here we first examine the nucleus mass loss from photometry by transforming Equation (\ref{eq_RN}) into
 \begin{equation}
 \dot{M}_{\mathrm{N}} \left( t \right) = \frac{4}{3} \rho_{\mathrm{d}} \left(\frac{3 - \gamma}{4 - \gamma} \right) \left(\frac{a_{\max}^{4 - \gamma} - a_{\min}^{4 - \gamma}}{a_{\max}^{3 - \gamma} - a_{\min}^{3 - \gamma}}  \right) \dot{C}_{\mathrm{e}} \left(t \right).
 \label{eq_mloss}
 \end{equation}
 \noindent To obtain $\dot{C}_{\mathrm{e}}$, first, smoothing with 10 neighboring data points is performed to the derived $C_{\mathrm{e}}$ (shown in Figure \ref{fig:xs_2015D1}). During tests we found that if too few neighboring data points are used, artifacts will be formed from the scattered data around the downhill portion of the pre-perihelion dip. On the other hand, there is no significant improvement if more neighboring data points are included. Next, we take the difference between each time step. With Equation (\ref{eq_mloss}), we then obtain a mass loss rate at each time, as shown in Figure \ref{fig:mloss_rate}. 
 
 While optical depth effects would delay the apparent time of mass loss, this should set a reasonable time boundary. We can see that the most rapid mass loss rate occurred around perihelion, with $\dot{M}_{\mathrm{N}} \sim 10^{5}$ kg s$^{-1}$. This is consistent with Section \ref{sim_morph} that the post-perihelion tail was formed during this period in a quasi-impulsive manner. It is noteworthy that negative mass loss rates do not necessarily reflect genuine variation; they can be better explained by particles continuously drifting out of the photometric aperture or sublimating without adequate resupply from the nucleus. We obtain the total mass loss around perihelion to be $\sim$10$^{9}$ kg, the same order of magnitude as the original nucleus mass estimated previously.
 
 We can also investigate the mass loss according to the non-gravitational effect. Equation (\ref{eq_mass_nuc}) can be transformed into an ordinary differential equation
 \begin{equation}
 M_{\mathrm{N}} \left(t \right) = -\kappa \frac{\dot{M}_{\mathrm{N}} \left(t \right) v\left(t \right)}{ \mathfrak{A}\left(t \right)} ,
 \label{eq_mass_loss}
 \end{equation}
 \noindent where $\dot{M}_{\mathrm{N}} = \mu Q$. The variables are separable and integrable. We can then solve the ratio of mass loss to the initial nucleus mass, $\mathcal{E}_{\mathrm{M}}$, during an observation interval from $t_{0}$ to $t_{\mathrm{obs}}$, by
 \begin{equation}
 \mathcal{E}_{\mathrm{M}} = 1 - \exp \left[ -\frac{\mathcal{A}}{\kappa} \int_{t_{0}}^{t_{\mathrm{obs}}} \frac{g\left(r_{\mathrm{h}} \left(t\right) \right)}{v\left(r_{\mathrm{h}} \left( t \right) \right) } \mathrm{d} t \right] .
 \label{eq_mass_loss2}
 \end{equation}
 \noindent Here we choose $t_{0}$ to be the time when the first LASCO observation of 2015 D1 was made, and $t$ varies from the beginning to the end of LASCO observation. We find that if the empirical water-ice sublimation model is correct, 93.2\% of the nucleus mass would have been lost by perihelion, and 99.6\% would have been eroded by the time 2015 D1 exited LASCO C3's FOV. While we agree that mass erosion is predominantly important to 2015 D1, judging from the photometric and morphological analysis, we suspect that this overestimates the mass loss by perihelion. 
 
 We then investigate the mass loss due to sublimation of forsterite and follow the same procedures described by Sekanina \& Kracht (2015) to calculate $\mathcal{A}_{1, \mathrm{for}}$ ($\mathcal{A}_{j, \mathrm{for}} = 0$ assumed for $j \ne 1$). The sodium model is skipped as we believe that the sodium amount is very small and therefore a significant mass loss due to sublimation of sodium is highly unlikely. During tests we found that the sodium sublimation model gives results very similar to those by water-ice sublimation. 
 
\begin{figure}
\begin{center}
\includegraphics[scale=1]{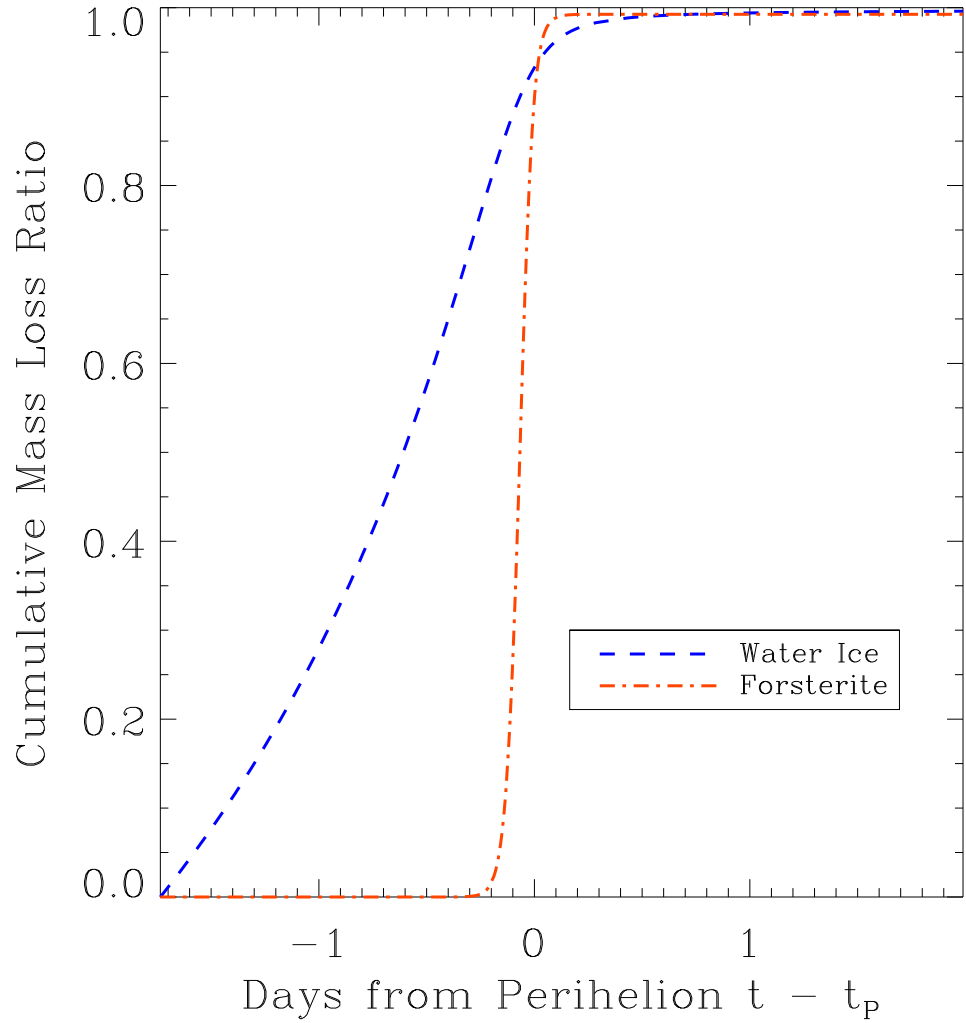}\\
\caption{
Modeled cumulative mass loss ratios from two different non-gravitational momentum transfer laws, i.e. water-ice and forsterite sublimation. The models are labeled on the plot and detailed discussions are in Section \ref{disc_mloss}.
\label{fig:mod_mloss}
}
\end{center}
\end{figure}

 We obtain $\mathcal{A}_{1, \mathrm{for}} = \left(3.990 \pm 0.362 \right) \times 10^{-33}$ AU day$^{-2}$. Comparisons between different models are shown in Figure \ref{fig:mod_mloss}, from which we can see that sublimation of forsterite would lead to the comet experiencing rocketing mass erosion once it reached a very small heliocentric distance of $r_{\mathrm{h}} \lesssim 8~R_{\odot}$. By perihelion, an overwhelmingly large section of the initial mass, 89.0\%, would be eroded, and the comet would devastatingly lose 99.3\% of the mass by the end of the LASCO observation. 
 
 However, the forsterite model exaggerates the sinusoidal envelope of the residuals in declination pre-perihelion (Figure \ref{fig:orb_resid}c) and slightly worsens the residuals, RMS $= \pm$11\arcsec.60, in comparison to the water-ice model. Most importantly, we find that sublimation of forsterite alone fails to support the enormous mass loss experienced by 2015 D1 around perihelion. To verify this, we apply equations and parameters in Kimura et al. (2002) to estimate $\dot{M}_{\mathrm{N}}$ due to sublimation of forsterite around perihelion. We obtain the unit area mass loss rate as $5.7\times10^{-7}$ kg s$^{-1}$ m$^{-2}$, which is then multiplied by the surface area of the nucleus, yielding $\dot{M}_{\mathrm{N}} \sim 7 \times 10^{-2}$ kg s$^{-1}$. Although a porous nucleus would increase the surface area, resulting in a larger $\dot{M}_{\mathrm{N}}$, yet it is still far too small compared to the mass loss of 2015 D1 around perihelion. In comparison, given an isothermal nucleus, the mass production rate of water-ice around perihelion is $\sim$0.2 kg s$^{-1}$ m$^{-2}$. In order to support the observed peak mass loss rate, this would require a surface area of $\sim$1 km$^{2}$, equivalent to a $\sim$0.3 km radius sphere. This is order-of-magnitude consistent with our estimate of the nucleus size before disintegration. Thus, the lack of other better models makes the isothermal water-ice sublimation model the best choice for approximation.

  \subsection{Mechanism of the Disintegration}
  \label{disc_breakup}
We briefly investigate the breakup of 2015 D1 since it is the only sunskirting comet for which there is strong observational evidence for fragmentation while it was being observed. We first consider the possibility that breakup was caused by tidal disruption due to its proximity to the Sun. For a non-spinning fluid body, the Roche radius of the Sun is $\sim$3.7 $R_{\odot}$ for a bulk density of $\rho_{\mathrm{d}} = 0.4$ g cm$^{-3}$, whereas 2015 D1 started to fragment at $r_{\mathrm{h}} \ge 6.06~R_{\odot}$. A comet experiencing tidal disruption at a distance $\sim$1.6 times larger than the Roche radius seems farfetched, although we cannot fully rule out the possibility because of unknown factors such as the nucleus density and the way the nucleus spins can affect the actual Roche radius (Asphaug \& Benz 1996, Richardson et al. 1998). 
    
  We next consider effects from thermal fracture. The timescale of heat conduction from the surface to the interior of a spherical rocky body is roughly $\tau_{\mathrm{H}} \sim R_{\mathrm{N}}^{2} / \kappa_{\mathrm{eff}}$ where $\kappa_{\mathrm{eff}} \sim 10^{-6}$ m$^{2}$ s$^{-1}$ is the effective thermal diffusivity typical for rocks. For 2015 D1, we have $\tau_{\mathrm{H}} \sim 10^{2}$ yr.
  
  The core temperature of 2015 D1 can be estimated from conservation of energy 
  \begin{equation}
  T_{\mathrm{C}} = \left[\frac{\left( 1 - A_{\mathrm{p}} \right) S_{\odot}}{4 \epsilon \sigma \left(t_{\mathrm{P}} - t_{0} \right)} \int_{t_{0}}^{t_{P}} \frac{\mathrm{d} t}{r_{\mathrm{h}}^{2} \left(t \right)}  \right]^{1/4} .
  \label{eq_T_core}
  \end{equation}
  \noindent We choose $t_{\mathrm{P}} - t_{0} = 100$ yr, and thereby obtain $T_{\mathrm{C}} \simeq 90$ K for the nucleus core. On the contrary around perihelion, with much of its nucleus surface devoid of volatiles assumed and sublimation of forsterite taken into account, the equilibrium surface temperature was $\sim$1640 K; a huge temperature gradient of $\Delta T \simeq 1550$ K from the nucleus surface to the interior would be formed. To estimate the established thermal stress we set a nominal thermal expansion coefficient, $\alpha_{\mathrm{V}} \sim 10^{-5}\mbox{--}10^{-6}$ K$^{-1}$, typical for common rocks, and a Young's modulus $Y \sim 10^{9}\mbox{--}10^{11}$ Pa (e.g., Jewitt \& Li 2010, Sekanina \& Chodas 2012, and citations therein). An overwhelmingly huge thermal stress, $\sigma_{\mathrm{th}} = \alpha_{\mathrm{V}} Y \Delta T \sim 10^{6}\mbox{--}10^{9}$ Pa would be generated inside its interior, which is an order of magnitude or more larger than typical tensile strengths of cometary nuclei (c.f. Prialnik et al. 2004 and citations therein). Thermal fracture and cracking were very likely to occur, whereby preexisting subsurface volatiles were exposed, disastrously intensifying the outgassing activity. 
  
  Recently, Steckloff et al. (2015) argued that differential stress within the nucleus interior due to dynamic sublimation pressure may have been responsible for the breakup of sungrazing comet C/2012 S1 (ISON). This mechanism might be plausible for C/2012 S1 (ISON) due to the fact that its nucleus had withstood strong outgassing activity for a long period of time ($\gtrsim 1$ yr) before disintegration, however, there is no evidence that 2015 D1 was similarly active. Thus, we favor thermal fracture for 2015 D1, since sublimation stress is likely orders of magnitude smaller than the thermal stress built up within the interior. 
   
  Yet it is still unclear whether the explosion of outgassing directly crumbled the nucleus. Even if not, the fate of the nucleus was destined not to survive. Torques exerted by large mass loss from the nucleus can lead to rotational instability. Observations suggest that rotational breakup is a very common fate for comets in the solar system (e.g., Jewitt et al. 1997). Concentrating around the perihelion passage, we justify this hypothesis by
  \begin{equation}
  \Delta R_{\mathrm{N}} \simeq \frac{4\pi R_{\mathrm{N}}^{2}}{15 \kappa_{\mathrm{T}} P_{\mathrm{rot}} v_{\mathrm{th}}} .
  \label{eq_rotinst}
  \end{equation}
\noindent which is derived in Li \& Jewitt (2015). Here $\kappa_{\mathrm{T}} \sim10^{-4}\mbox{--}10^{-2}$ is a dimensionless coefficient of the torque (e.g., Belton et al. 2011, Drahus et al. 2011), and $P_{\mathrm{rot}}$ is the rotation period of the nucleus. We assume a rotation period $P_{\mathrm{rot}} \sim 10^{5}$ s, which is typical for cometary nuclei (Samarasinha et al. 2004). The thermal speed around the perihelion is $v_{\mathrm{th}} \sim 1$ km s$^{-1}$. By substituting other numbers we obtain $\Delta R_{\mathrm{N}} \sim$ 0.01--1 m. Combined with Equation (\ref{eq_RN}), we find that around perihelion, it would take the nucleus an extremely short time, $\Delta t = \Delta R_{\mathrm{N}} / \dot{R}_{\mathrm{N}} \sim$ 1--100 s, to achieve such a change in the nucleus radius, which means that within such a short period of time, mass shedding due to outgassing would change the angular momentum by a significant factor. We hence see rotational instability as a plausible mechanism for a final blow to the nucleus by disintegrating it, provided that it survived the outgassing explosion. This agrees with Samarasinha \& Mueller (2013) that rotational disruption is likely the most common cause for splitting of sub-kilometer sized near-Sun comets. Note that rotational disruption is not significant at large heliocentric distances because the outgassing activity is limited.

\section{SUMMARY}

We present analysis of sunskirting comet C/2015 D1 (SOHO) using observations from \textit{SOHO}, Xingming, and Lowell. We conclude:

\begin{enumerate}

\item Non-gravitational effects experienced by this comet were obvious. Solving $\mathcal{A}_{1}$ in the orbital solution improves O$-$C residuals significantly and helps remove the sinusoidal trends. We find $\mathcal{A}_{1} = \left(1.209 \pm 0.118 \right) \times 10^{-6}$ AU day$^{-2}$, based upon the isothermal water-ice sublimation model. The non-gravitational acceleration was unlikely due to forsterite sublimation as there is insufficient sublimation to drive the observed mass loss around perihelion.

\item Photometric data and non-gravitational effects consistently suggest the pre-disintegration nucleus mass as $M_{\mathrm{N}} \sim 10^{8}\mbox{--}10^{9}$ kg, and the nucleus size as $R_{\mathrm{N}} \sim 50\mbox{--}150$ m in radius, with $\rho_{\mathrm{d}} = 0.4$ g cm$^{-3}$ assumed.

\item The mass loss was predominantly concentrated around its perihelion passage, with the most rapid loss as large as $\dot{M}_{\mathrm{N}} \sim 10^{5}$ kg s$^{-1}$. A significant portion of the mass was shed during this time interval, comparable to the original nucleus mass.

\item Morphological simulation of the comet's post-perihelion tail indicates that it was formed around $\Delta t \sim -1$ hr (UT 2015 February 19.7, or DOY $\sim$50.7) within 0.1 day, in a quasi-impulsive manner, when the comet suffered from the most rapid mass loss. The remnant of the debris cloud was morphologically dominated by smaller dust particles. The freshest dust grain sizes were $a \gtrsim 10~\mu$m, with $\rho_{\mathrm{d}} = 0.4$ g cm$^{-3}$ assumed. An increasing trend in $a_{\min}$ was noticed, which is likely due to the smaller dust grains being dispersed more quickly without further replenishment of dust and hence dimming gradually beyond the detection threshold. We thus derive a power law index $\gamma = 3.2 \pm 0.1$ for the dust size distribution.

\item We suggest that the flareup in brightness was likely triggered by excess thermal stress built up within the nucleus interior causing an explosive release of material and exposing subsurface volatiles. The outgassing explosion may have crumbled the nucleus. Even if not, subsequent rotational instability of the nucleus could easily lead to its disintegration. It would only take the nucleus a very short period of time, $\Delta t \sim $1--100 s, to change its angular momentum by a large factor.

\item The huge dip in the light curve starting from $\sim$8 hrs prior to perihelion is not due to the Swings effect. Mild turnover points at $r_{\mathrm{h}} \sim 13~R_{\odot}$ and the more obvious one at $r_{\mathrm{h}} \sim 8~R_{\odot}$ suggest that sublimation of olivines is likely responsible, which is directly supported by the disappearance of the pre-perihelion tail around the same time. The subsequent rapid brightening resulted from disintegration of its nucleus, which drastically increased the effective cross-section area.

\item The comet had a color distinctly different from the color of the Sun, in particular pre-perihelion, but gradually evolved to the solar color. Sodium content and not thermal emission was the most likely the cause of the color. Depletion of sodium emission led to a final color similar to that of the Sun, which implies that the nucleus exhausted its volatiles and the coma turned dusty.

\item Ground-based observations from Xingming and Lowell revealed no detectable central condensation in the debris cloud 13--24 days after perihelion. The post-perihelion non-detection from Lowell Observatory restricts any remaining active nucleus size to be $R_{\mathrm{N}} \lesssim 0.1$ km.

\end{enumerate}

\acknowledgments
Thanks to the anonymous referee for the prompt reviews and comments. We thank Aldo Vitagliano for providing us with his modified version of EXORB and thus enabling our need to investigate non-gravitational effects based upon different models; David Jewitt and Jing Li for reading the manuscript and providing comments; Rainer Kracht and Gaeul Song for their insightful and valuable discussions; Kevin Schenk for providing detailed technical information of \textit{SOHO}; and the \textit{SOHO} operation team for making their data publicly available. We also thank Xing Gao for scheduling observations and acquiring the CSP images, Teznie Pugh for obtaining the DCT images, Alberto Bolatto for providing DCT calibration images, and Ed Anderson for obtain the Lowell 31-in images. This research has made use of the VizieR catalogue access tool, CDS, Strasbourg, France, facilities of the Canadian Astronomy Data Centre operated by the National Research Council of Canada with the support of the Canadian Space Agency, SSW/SSWDB, IDL Astronomy User's Library and Bill Gray's Charon. The original description of the VizieR service was published in A\&AS 143, 23. M.-T.H. was supported by a NASA grant to David Jewitt. M.M.K. was supported by NASA Planetary Astronomy grant NNX14AG81G. K.B. was supported by the NASA-funded Sungrazer Project.

\begin{deluxetable}{lcrrrrrrrrr}
\tablecaption{Viewing Geometry of Ground Observations \label{geometry}}
\tablewidth{0pt}
\tablehead{ \colhead{Date (UT)} & Tel\tablenotemark{a}   & \colhead{$r_{\mathrm{h}}$\tablenotemark{b}}  & \colhead{$\Delta$\tablenotemark{c}} & \colhead{$\alpha$\tablenotemark{d}}  & \colhead{$\varepsilon$\tablenotemark{e}}  & \colhead{$\theta$\tablenotemark{f}} &   \colhead{PsAng\tablenotemark{g}}  & \colhead{PsAMV\tablenotemark{h}}   & \colhead{$X$\tablenotemark{i}}}
\startdata

2015-Mar-04 13:13-13:55 & CSP  & 0.579 & 0.869 & 83.8 & 35.5 & 154.1 & 38.8 & 206.3 & 2.73-4.07\\
2015-Mar-05 02:20-02:26 & L31  & 0.596 & 0.875 & 82.4 & 36.6 & 154.5 & 38.9 & 206.5 & 2.66-2.81\\
2015-Mar-05 02:48-02:51 & DCT &  0.596 & 0.876 & 82.3 & 36.6 & 154.5 & 38.9 & 206.5 & 3.59-3.67\\
2015-Mar-08 13:15-14:26 & CSP & 0.700 & 0.928 & 73.6 & 42.6 & 156.4 & 39.9 & 208.2 & 2.08-3.44\\
2015-Mar-09 13:37-14:41\tablenotemark{\dagger} & CSP & 0.729 & 0.946 & 71.3 & 44.1 & 156.9 & 40.3 & 208.8 & 2.24-3.59\\
2015-Mar-15 13:30-15:29 & CSP & 0.893 & 1.075 & 59.8 & 51.0 & 159.0 & 43.7 & 212.6 & 1.72-3.37\\
\enddata

\tablenotetext{a}{Telescope: CSP = Xingming Observatory's 0.11-m refractor; DCT = Lowell Observatory's 4.3-m Discovery Channel Telescope; L31 = Lowell Observatory's 31-in (0.8-m) reflector}
\tablenotetext{b}{Heliocentric distance, in AU}
\tablenotetext{c}{Cometocentric distance to the observatory, in AU}
\tablenotetext{d}{Phase angle, in degrees}
\tablenotetext{e}{Solar elongation, in degrees}
\tablenotetext{f}{True anomaly, in degrees}
\tablenotetext{g}{Position angle of the extended Sun-to-comet radius vector in the plane of sky, in degrees}
\tablenotetext{h}{Position angle of the projected negative heliocentric velocity vector, in degrees}
\tablenotetext{i}{Air mass, dimensionless}
\tablenotetext{\dagger}{Six images taken later than 14:24 UT were partially obstructed, hence discarded.}
\tablecomments{This table is compiled from JPL HORIZONS. We are aware that discrepancies between predicted positions increased over time. The worst case is for the CSP observation on 2015 March 15, where the JPL ephemeris differs from the one based upon EXORB by $\sim$0\degr.8. }

\end{deluxetable}

\clearpage

\begin{deluxetable}{cccccc}
\tablecaption{Orbital Elements
(Reference: Heliocentric Ecliptic J2000.0)
\label{orbit}}
\tablewidth{0pt}
\tablehead{ \colhead{Element} & \colhead{Value without $\mathcal{A}$} & \colhead{1$\sigma$ Uncertainty} & \colhead{Value with $\mathcal{A}$} & \colhead{1$\sigma$ Uncertainty} & \colhead{Units}
}
\startdata

$t_{\mathrm{P}}$\tablenotemark{a} & 2015 Feb 19.74859 & $5.3\times10^{-5}$ & 2015 Feb 19.74642 & $2.2\times10^{-4}$ & TT \\
$q$\tablenotemark{b} &  0.0284511 & $3.1\times10^{-6}$ & 0.028219 & $2.3\times10^{-5}$ & AU \\
$e$\tablenotemark{c} & 1.00099 & $2.1\times10^{-4}$ &1.00142 & $2.1\times10^{-4}$ & \\
$i$\tablenotemark{d} & 69.355 & $1.8\times10^{-2}$ & 69.582 & $2.9\times10^{-2}$ & deg \\
$\Omega$\tablenotemark{e} & 95.924 & $1.6\times10^{-2}$ & 95.897 & $1.7\times10^{-2}$ & deg \\
$\omega$\tablenotemark{f} & 235.194 & $9.4\times10^{-3}$ & 235.635 & $4.4\times10^{-2}$ & deg \\
$\mathcal{A}_1$\tablenotemark{g} & -- & -- & $+1.209\times10^{-6}$ & $1.18\times10^{-7}$ & AU day$^{-2}$ \\

\enddata
\tablenotetext{a}{~Time of perihelion passage in Terrestrial Time (TT)}
\tablenotetext{b}{~Perihelion distance}
\tablenotetext{c}{~Eccentricity}
\tablenotetext{d}{~Inclination}
\tablenotetext{e}{~Longitude of ascending node}
\tablenotetext{f}{~Argument of perihelion}
\tablenotetext{g}{~Water-ice sublimation model by Marsden et al. (1973). Only $\mathcal{A}_{1}$ solved, presumably $\mathcal{A}_{2} = \mathcal{A}_{3} = 0$.}
\tablecomments{The RMSs of the orbital solutions without and with $\mathcal{A}_{1}$ are $\pm$13\arcsec.37~and $\pm$10\arcsec.44, respectively. Both solutions have epochs on TT 2015 February 18.00483.}
\end{deluxetable}
\clearpage

\clearpage

\clearpage

\end{document}